\pdfoutput=1
\documentclass[a4paper,11pt]{article}
\usepackage{amsmath,amssymb,amsfonts}
\usepackage[multiple,stable]{footmisc}
\usepackage[amsmath,amsthm,thmmarks]{ntheorem}
\allowdisplaybreaks[4]
\usepackage[noconfig]{refstyle}
\usepackage[dvipsnames]{xcolor}
\usepackage[hyperfootnotes=false, linktocpage=true, colorlinks, citecolor=blue, linkcolor=blue, urlcolor=Maroon]{hyperref}
\usepackage{array,tabularx,booktabs,geometry,subfigure,graphicx,longtable}
\usepackage[bf]{caption}
\usepackage{tikz}
\usepackage{cite}
\usepackage{dsfont}
\usepackage{relsize}
\usetikzlibrary{shapes,arrows}
\tikzstyle{block} = [rectangle, draw, text width=7em, text centered, rounded corners, minimum height=3em]

\let\eqref=\relax
\newref{eq}{name={eq.~},Name={Eq.~},names={eqs.~},Names={Eqs.~},rngtxt={-},refcmd=(\ref{#1})}
\newref{f}{name={footnote~},Name={Footnote~},names={footnotes~},Names={Footnotes~}}
\newref{app}{name={appendix~},Name={Appendix~},names={appendixes~},Names={Appendixes~}}
\newref{tab}{name={table~},Name={Table~},names={tables~},Names={Tables~}}
\newref{ch}{name={chapter~},Name={Chapter~},names={chapters~},Names={Chapters~}}
\newref{sec}{name={section~},Name={Section~},names={sections~},Names={Sections~}}
\newref{fig}{name={figure~},Name={Figure~},names={figures~},Names={Figures~}}
\numberwithin{equation}{section}
\newcommand{\be}{\begin{equation}}
\newcommand{\ee}{\end{equation}}
\newcommand{\bea}{\begin{equation}\begin{aligned}}	% note: abbreviations for \begin{align} and \end{align} don't work!
\newcommand{\eea}{\end{aligned}\end{equation}}		% note: \begin{equation}\begin{split}... produces pdf/hyperref warnings:
\newcommand{\iddots}{\mathinner{\mkern2mu\raise1pt\hbox{.}\mkern2mu \raise4pt\hbox{.}\mkern2mu\raise7pt\hbox{.}\mkern1mu}}

\providecommand{\id}{\leavevmode\hbox{\small$\mathrm{1}$\kern-3.8pt\normalsize$\mathrm{1}$}}
\def\fnote#1#2{\begingroup\def\thefootnote{#1}\footnote{#2}
     \addtocounter{footnote}{-1}\endgroup}
\begin{document}

\vspace{1cm}

\title{
       {\Large \bf Free 	Quotients of Favorable Calabi-Yau Manifolds}}

\vspace{2cm}

\author{
James~Gray${}^{1}$
and
Juntao~Wang${}^{2,3}$
}
\date{}
\maketitle
\begin{center} {\small ${}^1${\it Department of Physics, 
Robeson Hall, Virginia Tech \\ Blacksburg, VA 24061, U.S.A.}\\[0.2cm]
${}^2${\it Yanqi Lake Beijing Institute of Mathematical Sciences and Applications,\\ Yanqihu, Huairou District, Beijing, 101408, China}\\[0.2cm]
${}^3${\it Yau Mathematical Sciences Center, Tsinghua University,\\ Beijing, 100084, China}
}\\

\fnote{}{jamesgray@vt.edu}
\fnote{}{juntao.wang@bimsa.cn}

\end{center}

\begin{abstract}
\noindent
Non-simply connected Calabi-Yau threefolds play a central role in the study of string compactifications. Such manifolds are usually described by quotienting a simply connected Calabi-Yau variety by a freely acting discrete symmetry. For the Calabi-Yau threefolds described as complete intersections in products of projective spaces, a classification of such symmetries descending from linear actions on the ambient spaces of the varieties has been given in \cite{Braun:2010vc}. However, which symmetries can be described in this manner depends upon the description that is being used to represent the manifold. In \cite{Anderson:2017aux} new, favorable, descriptions were given of this data set of Calabi-Yau threefolds. In this paper, we perform a classification of cyclic symmetries that descend from linear actions on the ambient spaces of these new favorable descriptions. We present a list of 129 symmetries/non-simply connected Calabi-Yau threefolds. Of these, at least 33, and potentially many more, are topologically new varieties.
\end{abstract}

\thispagestyle{empty}
\setcounter{page}{0}
\newpage

\tableofcontents

%%%%%%%%%%%%%%%%%%%%%%%%%%%%%%%%%%%%%%%%%%%%%%%%%%

\section{Introduction}\label{favcicysec}

Non-simply connected Calabi-Yau threefolds play a central role in many studies of string theory compactifications. In heterotic string model building, a non-trivial fundamental group is needed to allow for the possibility of Wilson line breaking of Grand Unified groups \cite{Candelas:1985en,Green:1987mn} (see \cite{Braun:2005ux,Bouchard:2005ag,Braun:2005nv,Anderson:2009mh,Anderson:2011ns,Braun:2011ni,Anderson:2012yf,Larfors:2020weh,Constantin:2021for} for recent examples with the exact spectrum of the minimal supersymmetric standard model). This is especially important given work which shows that other approaches to achieving the standard model gauge group are hard to realize \cite{Anderson:2014hia}. In many other applications, non-simply connected Calabi-Yau threefolds are used simply because, due to their construction via quotients, they often have fewer moduli. This leads to more tractable computations in providing concrete examples of various physical phenomena (see \cite{Anderson:2013qca,Anderson:2011ty} for some illustrative examples of this type). Unfortunately, non-simply connected Calabi-Yau threefolds are rare in the known set of such manifolds. A set of $1695$ such spaces associated to the complete intersection Calabi-Yau manifolds in products of projective spaces (CICYs) was initially studied in \cite{Candelas:2008wb} with a complete classification being provided in \cite{Braun:2010vc}. There are $16$ cases with non-trivial fundamental group in the set of hypersurfaces in toric varieties \cite{Batyrev:2005jc} and two more cases related to this data set were isolated in \cite{Braun:2017juz}. These are the only known examples related to these two constructions, which are the data sets most commonly used in the physics literature.

In this paper we will classify a new set of non-simply connected Calabi-Yau threefolds. This will be achieved by classifying a set of discrete, freely acting symmetries on simply connected Calabi-Yau manifolds, which can then be used to form a non-simply connected and smooth Calabi-Yau quotient. The symmetries we have classified are available at \href{http://www1.phys.vt.edu/~grayphys/CyclicSyms.html}{this} online database.

\vspace{0.2cm}

In \cite{Braun:2010vc}, Volker Braun classified a particular set of freely acting symmetries on the CICYs which give rise to smooth non-simply connected Calabi-Yau manifolds upon quotienting. The CICYs were first introduced and classified in \cite{Yau:1986gu,Hubsch:1986ny,Candelas:1987kf,Green:1986ck,Candelas:1987du}. These threefolds can be described in terms of a configuration matrix.
\begin{eqnarray} \label{config}
X=\left [ \begin{array}{c|cccc}
\mathbb{P}^{n_1} & q^1_1 & q^1_2 & \dots & q^1_k  \\
\mathbb{P}^{n_2} & q^2_1 & q^2_2 & \dots & q^2_k\\
\vdots & \vdots & \vdots & \ddots & \vdots  \\
\mathbb{P}^{n_m} & q^{m}_1 & q^{m}_2 & \dots & q^{m}_k
\end{array}
\right ]
\end{eqnarray}
Here, the left most column indicates that the Calabi-Yau manifold is being embedded in an ambient space $A= \mathbb{P}^{n_1}\times\mathbb{P}^{n_2}\times\dots\times\mathbb{P}^{n_m}$. The remaining columns of (\ref{config}) describe a set of $k$ homogeneous polynomial equations, or defining relations, in the coordinates of the ambient space, the common solution set to which is the Calabi-Yau threefold. The $q^i_j$ are integers which specify the degree of the $j$'th defining relation in terms of the homogeneous coordinates of the $i$'th ambient space projective factor. In \cite{Candelas:1987kf} a classification of these manifolds was given. A list of $7890$ configuration matrices was provided (from here on termed the `original CICY list') such that any manifold of this type is described by at least one of those matrices. All of the CICY threefolds are simply connected. 

The set of symmetries of the CICYs classified in \cite{Braun:2010vc} were those which descended to the Calabi-Yau from linear actions on the ambient space coordinates and defining polynomials (a more precise definition will be given in Section \ref{bigscansec}). Such symmetry actions are relatively easy to control computationally, but they are not expected to be exhaustive in any sense. In particular, it is clear that which such symmetries are manifested will depend upon the configuration matrix. That is, the ambient space and polynomial equations being used to describe the Calabi-Yau threefold. The same manifold can be described by many different configuration matrices, both matrices within the original CICY list itself and matrices which are not included therein. One would therefore expect that some extra symmetries of these Calabi-Yau manifolds could be uncovered by studying different configuration matrices than those presented in the original list.

One particular type of configuration matrix turns out to be particularly useful in pursuing concrete computations of various geometrical quantities. A CICY $X$ is termed as `favorable' if $H^{1,1}(X)$ is spanned by a basis of K\"ahler forms pulled back from the $\mathbb{P}^n$ factors of the ambient space ${A}$. In many physical applications, a complete knowledge of divisor classes on $X$ is required in order to sufficiently control the geometry to compute the quantities of interest. Favorable manifolds are typically used in such situations and, for example, the majority of the example applications cited above were based upon threefolds of this type. It should be noted that, frequently, properties of a Calabi-Yau quotient are computed using the geometry of its covering space. Thus we are often interested in cases where the covering space, as well as the quotient, are favorable.

For the configuration matrices given in the original CICY list \cite{Candelas:1987kf,cicylist} only 4896 out of the 7890 configuration matrices are favorable. This restricts the set of manifolds that are usable in many applications. In \cite{Anderson:2017aux}, favorable configurations were found for all but 48 of the non-favorable CICYs presented in the original list. This was achieved using the concept of `splitting' of configuration matrices \cite{Candelas:1989ug}.

Consider starting with some configuration matrix with ambient space $A$. In the configuration matrix we isolate one column $c$ which will play the central role and denote the rest of the columns, which will spectate through the splitting process, by the matrix $C$. One can then split the original configuration matrix to obtain a new one as follows.
\begin{eqnarray}\label{splitting}
\left[
\begin{array}{c|cc}
 A & c & C
\end{array}
\right]\longrightarrow
\left[
\begin{array}{c|ccccc}
\mathbb{P}^n & 1 & 1 & \dots & 1 & 0\\
A & c_1 & c_2 & \dots & c_{n+1} & C
\end{array}
\right]
\end{eqnarray}
Here the columns $c_i$ must be chosen to obey the condition $c=c_1+c_2+\dots+c_{n+1}$. As was shown in \cite{Candelas:1989ug}, in general there is a conifold transition between the manifolds described by the two configuration matrices given in (\ref{splitting}), with the common point in moduli space being a nodal variety with some number of singular points. The number of these singular points can be determined by the difference in Euler characteristic between the two varieties. In particular, if the Euler characteristic of the two CICYs is the same, then the common point in moduli space is actually smooth and thus the manifolds have the same topology: they are different descriptions of the same CICY. Such a split is called ineffective.

In \cite{Anderson:2017aux}, chains of ineffective splits were performed to generate favorable configuration matrices for CICYs that are unfavorable in the original list. The essential idea is that, if the splits are performed in the correct manner, because there are more ambient factors after a split, additional divisors can descend from the ambient space. Sufficiently many splits can result in a favorable configuration matrix. In what follows, we will refer to the resulting list of configuration matrices as the `favorable CICY list'. For the 48 CICYs for which no favorable configuration was given, it was shown that no chain of splittings can give rise to such a description (alternative ways of controlling the divisors on $X$ were given in those cases). It should be noted that in \cite{Anderson:2017aux} just one favorable configuration was isolated for each CICY and others certainly exist. This is a point that we shall return to in Section \ref{resultssec}.

\vspace{0.3cm}

The freely acting symmetries, and thus non-simply connected Calabi-Yau manifolds, that we will classify in this paper are those cyclic symmetries which descend from the ambient spaces of the configuration matrices found in the favorable CICY list. We will see that some symmetries on some manifolds that were found in \cite{Braun:2010vc}, descending from the ambient space of the original CICY list, are not manifest in the symmetries that descend from the ambient space of the favorable CICY list. Likewise, we will find some symmetries in the favorable description which were previously unknown. Thus, in addition to providing new non-simply connected threefolds, this investigation will give us a very rough first look in to how many symmetries may have been missed in focussing on a single description of each manifold in the original classification of \cite{Braun:2010vc}.

\vspace{0.3cm}

The rest of this paper is structured as follows. In Section \ref{bigscansec} we review the rather technical algorithm, developed in \cite{Braun:2010vc}, for classifying the symmetries associated to a given configuration matrix. In Section \ref{resultssec} we will give some statistics concerning the symmetries we have found and will also specify the format in which our results are presented in the \href{http://www1.phys.vt.edu/~grayphys/CyclicSyms.html}{database associated to this paper}. We will also make some remarks and provide some examples in comparing our results with those of \cite{Braun:2010vc}. Finally in Section \ref{concsec} we conclude and provide some suggestions for future directions of research. An appendix is provided containing some useful technical formulae which are required in implementing the classification algorithm.

\section{Scanning Algorithm} \label{bigscansec}

\subsection{Symmetries from linear actions on ambient spaces}

The set of discrete symmetry actions that we are going to study can be thought of as combinations of the following components:
\begin{itemize}
\item Projective linear actions on the homogeneous coordinates of the ambient $\mathbb{P}^{n_i}$;
\item Permutations of rows and columns of CICY configuration matrix;
\item Projective linear actions on the defining polynomials.
\end{itemize}

In fact, in this work, we will restrict ourselves to cyclic groups $\mathbb{Z}_n$. In such a context we can simply consider linear actions on the coordinates rather than projective linear ones. The reason for this is that any projective linear action can be turned into a linear one by the simple expedient of multiplying the matrices of the representation by constants. Formally, for the cyclic groups, the second group cohomology vanishes and the Schur cover of the group is the same as the group itself. Changing the representations matrices in this manner does not physically modify the symmetry, due to the scale invariance of the projective spaces and the invariance of the Calabi-Yau manifold under rescalings of the defining relations. 

\vspace{0.2cm}

\noindent {\bf The formal description of the group action}

\vspace{0.1cm}

We will now follow the procedure given in \cite{Braun:2010vc}, for formally describing symmetries of the type just discussed. Let us begin by considering how the action of the group $G$ of interest permutes the rows and columns of the configuration matrix. If $G$ is a symmetry group of a certain Calabi-Yau manifold $X$ with configuration matrix $C$, then for any $g\in G$, the form of $C$ should be invariant after $g$'s action on the rows and columns. To formalize this, Braun defined a quadruple $(C, G, \pi_r,\pi_c)$, with elements defined as follows.
\begin{itemize}
\item $C$ is the data of the configuration matrix of the Calabi-Yau manifold, which can be formally written as $(d_i, c_{ij}, \delta_{j})_{1\leq i\leq m, 1\leq j\leq k}$. Here $m$ and $k$ are integers which represent the number of projective factors in the ambient space and the number of distinct multi-degrees of defining polynomials of the CICY respectively. Each $d_i$ represents the number of homogeneous coordinates of the $i$-th projective space. The integers $c_{ij}$ give the degree of the $j$-th distinct column of the configuration matrix in terms of the homogeneous coordinates of the $i$'th ambient space projective factor. Finally, $\delta_j$ represents the multiplicity with which $j$-th distinct set of defining polynomial multi-degrees appears in the configuration matrix. 
\item $G$ is a finite order group which will act on the Calabi-Yau manifold.
\item $\pi_r:G\rightarrow P_{\text{row}}$ is a group homomorphism from $G$ to the set of permutations of elements of $\vec{d}$, whose entries are the $d_i$ defined above. It details how the group action permutes the rows of the configuration matrix.
\item $\pi_c:G\rightarrow P_{\text{col}}$ is a group homomorphism from $G$ to the set of permutations of elements of $\vec{\delta}$ defined above. It details how the group action permutes distinct columns of the configuration matrix.
\end{itemize}

 Given this notation, in order for the action of $G$ to give a symmetry of $X$, we must have that for every group element $g\in G$,
  \begin{eqnarray} \label{cinv}
c_{ij}=c_{\pi_r(g)(i) \pi_c(g)(j)}.
\end{eqnarray}
A quadruple $(C, G, \pi_r,\pi_c)$ which obeys these conditions was termed a CICY group in \cite{Braun:2010vc}.
It should be noted that once $C, G$ and $\pi_r$  have been given $\pi_c$ is determined uniquely by (\ref{cinv}). Obtaining this uniqueness is the reason why the configuration matrix data is presented in the above fashion, with multiple copies of same multi-degree defining relations being denoted by a single column of $c_{ij}$ and a multiplicity $\delta_j$. 
 
\vspace{0.3cm}

Now let us discuss how $G$ acts on the homogeneous coordinates of $X$ and also more detailed aspects of the action on the defining relations.  As we have seen above, a CICY group can tell us how $G$ can permute elements in the $\vec{d}$ and $\vec{\delta}$ vectors. Elements related by permutations in this manner form orbits of the action of $G$. To specify the action of $G$ on the coordinates and defining relations, one can then first specify the action on each orbit. The full action is subsequently simply given by a direct sum of each of these component pieces. 

\vspace{0.1cm}

Following \cite{Braun:2010vc} we will use the concept of a $\pi$-representation for which we reproduce the definition below. This concept will be used in detailing the transformation of both homogeneous coordinates and defining polynomials.

\vspace{0.1cm}

A $\pi$-representation is a quadruple $(G', \pi, \overrightarrow{D},\gamma)$ where
\begin{itemize}
\item $G'$ is a finite order group, usually we will have that $G'=G$.
\item $\vec{D}=(D_1,D_2,..,D_m)$ is an ordered set of numbers $D_i$. This $\vec{D}$ will be either $\vec{d}$ or $\vec{\delta}$ in our case.
\item $\pi:G\rightarrow P$ is a group homomorphism from $G$ to a permutation group $P$ acting on $\vec{D}$ defined above.
\item For each $D_i \in \vec{D}$, we have a map $\gamma_i: G\rightarrow GL(D_i, \mathbb{C})$. For any $g\in G$, $\gamma_i$ has the following property.
 \begin{eqnarray}\label{gammaconsnew}
\gamma_{\pi(h)(i)}(g)\gamma_{i}(h)=\gamma_i(gh)
\end{eqnarray}
\item Equation (\ref{gammaconsnew}) implies that all the $\gamma_i$'s can assembled into the the following matrix:
\begin{eqnarray}\label{gammabig}
\gamma(g)=P(\pi(g), \vec{D})\text{diag}(\gamma_1,\dots, \gamma_m).
\end{eqnarray}
Here $P(\pi(g), \vec{D})$ represents how $\pi(g)$ permutes the elements in $\vec{D}$ vector. 
\end{itemize}

\vspace{0.1cm}
 
Given this definition, a CICY group action is then defined to be a tuple $(C, G, \pi_r, \gamma, \pi_c, \rho)$ where the following statements hold.
\begin{itemize}
\item $(C, G, \pi_r, \pi_c)$ is a CICY group.
\item $(G,\pi_r, \vec{d}, \gamma)$ and $(G,\pi_c, \vec{\delta}, \rho)$ are $\pi$ representations.
\end{itemize}
The additional structure in the second bullet point above encapsulates the information that the CICY group action contains beyond that present in the associated CICY group. Specifically, the two $\pi$ representations account for the action on the homogeneous coordinates of the ambient space projective factors and also the action on the defining equations not captured by the permutations of distinct multi-degrees appearing in the data of the CICY group.

\vspace{0.3cm}

One key question that will arise in what follows is when a given set of defining relations for a CICY are compatible with a given symmetry. Let us denote the combined homogeneous coordinates of a CICY by $\vec{x}$ and the defining polynomials by $\vec{p}$. Then, we say that $\vec{p}$ is invariant under a CICY group action iff the following condition is satisfied.
\begin{eqnarray}\label{invarpoly}
\rho^{-1}(g)\vec{p}(\gamma(g)\vec{x})=\vec{p}(\vec{x})\quad \forall g\in G
\end{eqnarray}
If we fix a CICY group $(C, G, \pi_r, \pi_c)$ and a projective $\pi$-representation $(G,\pi_r, \vec{d}, \gamma)$ acting on the homogeneous coordinates, then the zero set of the defining polynomials $\vec{p}=0$ is invariant if an only if there is a CICY action $(C, G, \pi_r, \gamma, \pi_c, \rho)$ leaving the polynomials invariant in the above sense.

\vspace{0.3cm}

\noindent{\bf  Using orbits and characters to describe symmetries efficiently}

\vspace{0.1cm}

If one tries to enumerate the symmetries of the CICYs of the form discussed above by brute force, one quickly sees that the numbers of possibilities involved are prohibitively large. However, in \cite{Braun:2010vc}, two tools were used to circumvent this problem and make enumerating the symmetries on the CICYs practically feasible. First, although the number of possible CICY group actions is vast, they are built out of a smaller number of actions associated to individual orbits of the $\pi$-representations $(G,\pi_r, \vec{d}, \gamma)$. One can thus enumerate this smaller set of possibilities and then later construct the representations of interest as simple Cartesian products. Second, for many purposes, instead of specifying the detailed coordinate actions it is sufficient to encode the representations in terms of characters. In the next section, we will describe how these two tools are used in the search algorithm to classify the symmetries of interest. In this section, we simply introduce some of the technology necessary to follow such an approach.

\vspace{0.1cm}

As discussed above, for a given $\pi$ representation $(G,\pi_r, \vec{d}, \gamma)$, $\pi_r$ will break $\vec{d}$ into several orbits. The problem of finding representations of $G$ on a single orbit can be solved by using what is called induction, a process which we now describe. 

Clearly, $\pi_r$ can only permute $d_i$s which are equal to each other. Therefore $d_i$ will be a fixed number on each orbit which we will denote as $d$ below.  For a single orbit indexed by $\{1,\dots,n\}$, $G$ will permute those indices in a certain way. There are some group elements which act trivially on the first element in the orbit. These form the stabilizer of $1$ and we denote it as $G_1$:
\begin{eqnarray}\label{stabilizer1}
G_1=\{g\in G\;|\; \pi_r(g)(1)=1\}.
\end{eqnarray}
Let us denote by $\gamma_1$ the restriction of $\gamma$ to the first projective factor. For a representation $\gamma$ of $G$, because $G_1$ does not permute the first entry in our set at all, $\gamma_1$'s restriction on $G_1$ is a  representation of $G_1$. Once we know $\gamma_1|_{G_1}$, it turns out that we can recover the original representation $\gamma$ of $G$ on the whole orbit. We first fix $g_1=1$, then for $i=1,\dots, n$, we choose $g_i$ such that $\pi(g_i)(1)=i$. By using this set of group elements $\{g_1,\dots, g_n\}$, any group element g can be factorized as:
\begin{eqnarray}\label{factorize}
\forall g \in G, \quad\forall  \;1\leq i\leq n, \quad\exists h \in G_1: g=g_{\pi(g)(i)}\circ h\circ g^{-1}_i .
\end{eqnarray}
Since we choose $g_1=1$, we have that $\gamma_1(g_1)=\bold{1}_{d \times d}$. Since each $g_i$ will map the first block to the $i$-th block, we can, by a suitable choice of coordinates on the $i$'th block, choose $\gamma_1(g_i)=\bold{1}_{d\times d}$ as well. Now we can use (\ref{gammabig}) and (\ref{factorize}) to derive the following.
\begin{eqnarray}\label{factorizerep}
\gamma_i(g)=\gamma_1(g_{\pi(g)(i)})\gamma_1(h)\gamma_1(g_i)^{-1}=\gamma_1(h) \quad \forall i=1,2,\dots, n
\end{eqnarray}
Thus, given the representation $\gamma_1(h)$, and the permutations associated to the CICY group, we can recover the full action of the form (\ref{gammabig}) on our orbit. This procedure is called induction and $\gamma$ thus obtained is denoted by $\gamma=\text{Ind}^G_{G_1}\gamma_1$. So to summarize, we can enumerate $G$'s representations on a single orbit by enumerating one block's stabilizer's representations on that block, and then using induction.  

\vspace{0.3cm}

A second key piece of technology in what follows is the use of characters. For semi-simple representations, the character, a list of numbers where each element is the trace of the representation matrix associated to each group element, is in one-to-one correspondence with the representation itself up to isomorphism. Thus a lot of the information that we require can be encoded in terms of characters, which are computationally efficient to work with. In particular, the character of the induced representation $\text{Ind}^G_{G_1}\gamma_1$ can be efficiently obtained from the character of the representation of the stabilizer subgroup $\gamma_1$. There exists an inner product for characters, which is defined as:
\begin{eqnarray}\label{charinner}
\langle \chi, \psi\rangle=\frac{1}{|G|}\sum_{g\in G}\chi(g)\overline{\psi(g)}.
\end{eqnarray}
With respect to this inner product, for a group $G$ and its subgroup $G_1$, characters $\chi$ of $G_1$ and characters $\psi$ of $G$ will satisfy:
\begin{eqnarray}\label{inducedchar}
\langle \text{Ind}^G_{G_1}(\chi), \psi\rangle=\langle \chi, \textnormal{Res}^G_{G_1}(\psi)\rangle.
\end{eqnarray}
In this expression, if $\chi$ is the character of some representation $\gamma_1$ of $G_1$ then $\textnormal{Ind}^G_{G_1}(\chi)$ is the character of $\textnormal{Ind}^G_{G_1} \gamma_1$. In addition, $\textnormal{Res}^G_{G_1}(\psi)$ is the character obtained by restricting $\psi$ to the subgroup $G_1$. The data that can be obtained from this inner product is enough to reproduce $\textnormal{Ind}^G_{G_1}(\chi)$. In particular, simply evaluating (\ref{inducedchar}) multiple times, once taking $\psi$ to be the character of each irrep of $G$, is enough information to reconstruct the complete list of numbers $\textnormal{Ind}^G_{G_1}(\chi)$.

So in general, without constructing explicitly the induced representation, we can get the relevant characters associated to $G$ by using characters associated to $G_1$ from (\ref{inducedchar}). Therefore, by enumerating all of the characters of $G_1$, we can enumerate all the associated characters of $G$ by using induction. From here it is a simple task to obtain the characters of a full cartesian product group action.

\subsubsection{An example} \label{homogcoordeg}

As an illustrative example, let us take $G=\mathbb{Z}_4$ and consider CICY number 7300 in the standard list \cite{Candelas:1987kf,cicylist}, whose configuration matrix is given below. 
\begin{eqnarray} \label{cicy7300}
\left [ \begin{array}{c|ccc}
\mathbb{P}^{1} & 1 & 0 & 1  \\
\mathbb{P}^{1} & 0 & 1 & 1  \\
\mathbb{P}^{1} & 0 & 1 & 1  \\
\mathbb{P}^{1} & 0 & 1 & 1  \\
\mathbb{P}^{1} & 1 & 0 & 1  \\
\mathbb{P}^{1} & 1 & 0 & 1  \\
\end{array}
\right ].
\end{eqnarray}
In this case, since the configuration matrix, $C$, has no repeated equation degrees, the above matrix is also $c$. The ambient space of this CICY is $\mathbb{P}^1\times\mathbb{P}^1\times\mathbb{P}^1\times\mathbb{P}^1\times\mathbb{P}^1\times\mathbb{P}^1$. Since all of the projective spaces in this ambient space are $\mathbb{P}^1$'s and since all of the entries in each row are either $0$ or $1$, in principle all of the rows can be permuted with each other. Thus $\mathbb{Z}_4$ should act as a permutation on the set $\{1,2,3,4,5,6\}$. There are lots of choice for $P_{row}$ in such a case, but here we shall choose it to be generated by $q=(13)(26)(45)$, and we define $\pi_r$ to be induced by the map $g\rightarrow q$ where $g$ is the generator of $\mathbb{Z}_4$. This basically means that $g$ will exchange row $1$ and row $3$, row $2$ and row $6$, row $4$ and row $5$. Given this, $g^2$ induces no permutation on the rows and the effect of the action of $g^3$ will be the same as that of $g$. After $g$'s action on the rows, the CICY configuration matrix becomes the following.
\begin{eqnarray} \label{cicy73002}
\left [ \begin{array}{c|ccc}
\mathbb{P}^{1} & 0 & 1 & 1  \\
\mathbb{P}^{1} & 1 & 0 & 1  \\
\mathbb{P}^{1} & 1 & 0 & 1  \\
\mathbb{P}^{1} & 1 & 0 & 1  \\
\mathbb{P}^{1} & 0 & 1 & 1  \\
\mathbb{P}^{1} & 0 & 1 & 1  \\
\end{array}
\right ]
\end{eqnarray}
By comparison of (\ref{cicy7300}) and (\ref{cicy73002}), we can see that if we choose $\pi_c$ to be induced by $g\rightarrow (12)$, then $g$ will interchange column $1$ and column $2$ of the above configuration which will then be invariant under the action of $g$. Therefore, the $(C, G, \pi_r, \pi_c)$ we have chosen here form a CICY group as defined in the discussion around (\ref{cinv}). 

\vspace{0.3cm}

For the configuration matrix (\ref{cicy7300}), we have that $\vec{d}=(2,2,2,2,2,2)$ and $\vec{\delta}=(1,1,1)$.
As an example of $\pi$ representation, let us study the orbit $\{1,3\}$ in the above example. We are thus interested in a $G=\mathbb{Z}_4$ symmetry acting on $X=\mathbb{P}^1\times\mathbb{P}^1$. If we denote the elements of $G$ by $\{1,g,g^2,g^3\}$ then the stabilizer of $1$ in this example is the group $G_1=\mathbb{Z}_2$ associated to the set $\{1,g^2\}$. Following the discussion of the previous page we choose $g_1=1$ and $g_2=g$ and then choose coordinates such that $\gamma_1(g_1)=\gamma _1(1)  = \mathbb{I}_{2\times 2}$ and $\gamma_1(g_2)=\gamma _1(g)  = \mathbb{I}_{2\times 2}$.

Given this setup, we should be able to choose a representation of $G_1= \mathbb{Z}_2$ on the first $\mathbb{P}^1$ and from there induce a representation of $G=\mathbb{Z}_4$ on $\mathbb{P}^1 \times \mathbb{P}^1$. We will illustrate how this works in two different ways. First, we will work with the representations directly themselves. Second, we will work with characters to obtain the same information in a more computationally efficient, albeit more abstract, manner.

To choose our representation of $G_1$ on the homogeneous coordinates of the first $\mathbb{P}^1$ we simply need to specify $\gamma_1(g^2)$ given that the above conventions already specify $\gamma_1(1)$. For the purposes of this example we will choose,
\begin{eqnarray} \label{smallrep}
\gamma_1(g^2) =\left(\begin{array}{cc} 1 &0 \\ 0&-1 \end{array} \right)\;.
\end{eqnarray} 
With the specification of the representation of $G_1$ on $\mathbb{P}^1$ completed with this choice, we can now induce an action of $G$ on $\mathbb{P}^1 \times \mathbb{P}^1$. In detailing this larger representation, via (\ref{gammabig}), we can use the above discussion to observe that  $P(\pi(g), \vec{D})$ (where $\vec{D}=(2,2)$) will be given by the following.
\begin{eqnarray}\label{largep}
P(\pi(g), \vec{D})=\left(
\begin{array}{cc}
 0 & \mathbb{I}_{2\times 2} \\
 \mathbb{I}_{2\times 2} & 0 \\
\end{array}
\right)
\end{eqnarray}
Therefore, with the rest of the information given above, we find that we simply need to obtain expressions for $\gamma_1(g^3)$ and all of the $\gamma_2$'s. To do this, we use (\ref{factorize}) to obtain the appropriate $h\in G_1$ in each case and then (\ref{factorizerep}) to write down the relevant matrix, using our known quantities. We find the following.
\begin{eqnarray}
\gamma_1(g^3) = \left(\begin{array}{cc} 1&0\\0&-1 \end{array}\right) \;\;,\;\; \gamma_2(1) &=& \left(\begin{array}{cc} 1&0\\0&1 \end{array}\right) \;\;,\;\;\gamma_2(g) = \left(\begin{array}{cc} 1&0\\0&-1 \end{array}\right) \;\;,\;\;\\\nonumber
\gamma_2(g^2) = \left(\begin{array}{cc} 1&0\\0&-1 \end{array}\right) \;\;&,&\;\;\gamma_2(g^3) = \left(\begin{array}{cc} 1&0\\0&1 \end{array}\right) 
\end{eqnarray}

Given these quantities we can now use (\ref{gammabig}) to write down the induced representation on our orbit $\mathbb{P}^1 \times \mathbb{P}^1$. We find,
\begin{eqnarray} \label{fullrepeg}
\gamma(1) =\left(\begin{array}{cccc} 1&0&0&0 \\0&1&0&0 \\0&0&1&0 \\0&0&0&1 \end{array} \right) \;\;,\;\;\gamma(g) =\left(\begin{array}{cccc} 0&0&1&0 \\0&0&0&-1 \\1&0&0&0 \\0&1&0&0 \end{array} \right) \;\;,\;\;\\ \nonumber
\gamma(g^2) =\left(\begin{array}{cccc} 1&0&0&0 \\0&-1&0&0 \\0&0&1&0 \\0&0&0&-1 \end{array} \right) \;\;,\;\;\gamma(g^3) =\left(\begin{array}{cccc} 0&0&1&0 \\0&0&0&1 \\1&0&0&0 \\0&-1&0&0 \end{array} \right) \;\;.
\end{eqnarray}
It can be simply checked that the above is indeed a good representation of $\mathbb{Z}_4$ as expected.

A similar analysis could have been repeated for each of the three orbits in the row permutation action we are considering on (\ref{cicy7300}). There is in fact only one other (trivial) inequivalent choice we could have made in specifying the $G_1$ representation on the homogeneous coordinates of $\mathbb{P}^1$. Thus, given two choice for each of three orbits we can now easily understand eight different representations. In doing this we only need to explicitly analyze two possibilities on a smaller system $\mathbb{P}^1\times \mathbb{P}^1$, clearly leading to some computational advantage. This on its own will not be enough to make the problem of exhaustively analyzing symmetries tractable however, and thus we now repeat this computation using the technology of characters.

\vspace{0.2cm}

Instead of specifying the representation $\gamma_1$ of $G_1$ explicitly as above we could simply give its character. Taking the trace of $\gamma_1(1)$ and $\gamma_1(g^2)$ above we find that we would specify the character $\chi=(2,0)$ in this case. It is then simple to derive the character $\textnormal{Ind}_{G_1}^G(\chi)$ of the representation  (\ref{fullrepeg}) using (\ref{inducedchar}). The characters of the irreducible representations of $\mathbb{Z}_4$ are as follows.
\begin{eqnarray} \label{char1}
\psi_1=(1,1,1,1) \;,\; \psi_2= (1,i,-1,i) \;,\; \psi_3=(1,-1,-i,i) \;,\; \psi_4 = (1,-i,i,-1)
\end{eqnarray}
The restrictions of these to the subgroup $G_1 =\mathbb{Z}_2$ given above is then,
\begin{eqnarray} \label{char2}
\textnormal{Res}_{G_1}^G(\psi_1) = (1,1) \;,\;\textnormal{Res}_{G_1}^G(\psi_2) =(1,-1) \;,\; \textnormal{Res}_{G_1}^G(\psi_3) = (1,-i) \;,\; \textnormal{Res}_{G_1}^G(\psi_4) = (1,i) \;.
\end{eqnarray}
Using $\chi=(2,0)$ and (\ref{char1}) and (\ref{char2}) in (\ref{inducedchar}) we find the following.
\begin{eqnarray} \label{badger3}
\langle \textnormal{Ind}_{G_1}^G(\chi), \psi_1\rangle = \langle \textnormal{Ind}_{G_1}^G(\chi), \psi_2\rangle = \langle \textnormal{Ind}_{G_1}^G(\chi), \psi_3\rangle = \langle \textnormal{Ind}_{G_1}^G(\chi), \psi_4\rangle = 1
\end{eqnarray}
This is enough for us to identify $\textnormal{Ind}_{G_1}^G(\chi)=(4,0,0,0)$, which indeed is reproduced by the trace of the matrices in (\ref{fullrepeg}).

Thus, for applications where we only need to know the characters of the representations involved, one can simply specify the character of the representation given by $\gamma_1$ restricted to the subgroup $G_1$ (together with the relevant row permutations) and from there one can directly obtain the character of the representation induced on the full orbit. The representation itself does not need to be presented explicitly. For many, but not all, of the steps in the classification algorithm of \cite{Braun:2010vc}, knowledge of the relevant characters is sufficient, which facilitates computation greatly.

\subsection{The algorithm} \label{mralg}

In this subsection we present the algorithm, first developed in \cite{Braun:2010vc}, for classifying the CICY group actions on the CICYs. We will be applying this algorithm to the newly found favorable CICY list.
\begin{enumerate}
\item First we constrain which group orders could possibly appear for each CICY by evaluating certain topological indices which can be computed purely from the data given in the configuration matrix \cite{Candelas:1987du}. More precisely, the indices
\begin{eqnarray}
\chi({\cal N}^k \otimes TX^l) &=& \int_X \textnormal{td}(TX) \wedge \textnormal{ch}({\cal N}^k \otimes TX^l) \\ \nonumber
\sigma({\cal N}^k \otimes TX^l) &=& \int_X L(X) \wedge \tilde{\textnormal{ch}}({\cal N}^k \otimes TX^l)
\end{eqnarray}
must give an integer when divided by any group order of a possible freely acting symmetry (for any $k$ and $l$). Here ${\cal N}$ is the normal bundle and $TX$ the tangent bundle of the Calabi-Yau manifold. The quantities $\textnormal{td}$ and $\textnormal{ch}$ are the Todd and Chern characters as usual, $L$ is the Hirzebruch L-polynomial and $\tilde{\textnormal{ch}}$ the Chern character where the curvature form is replaced with twice its value. This is a computationally cheap initial step that leads to a finite number of possible group orders for each configuration matrix. Since there are only a finite number of finite groups of each order, this step is already sufficient to make the problem bounded (if still formidable).
\item For each possible symmetry order found above, we next obtain all the possible CICY groups, i.e., all the favorable configuration matrices' row and column permutations which leave the configuration matrix invariant. This step is possible to perform by simple brute force computation.
\item Next a necessary but not sufficient check for fixed point freeness is performed. This step, which will be described in detail below, only requires us to specify the character of the CICY group action, not the full $\pi$ representations themselves. This allows for the elimination of many possibilities before the full symmetry action has to be written down in detail, a crucial step in rendering the problem computationally tractable.
\item For each possible potential symmetry which has not been ruled out by the proceeding steps a necessary and sufficient check for fixed point freeness is performed by direct computation. This requires us to explicitly specify the CICY group action and invariant defining polynomials at this stage. The fixed point freeness is checked for each Calabi-Yau manifold starting from lower and proceeding to high order of the possible symmetries. If a symmetry group has non-fixed point free (or non-smooth, see the next point) sub-groups, then the full group can not be fixed point free (smooth) either and thus need not be checked.
\item Finally, for the fixed point free CICY group actions a standard Gr\"obner basis computation is used to ascertain whether the generic invariant defining polynomials lead to a smooth manifold. These calculations are performed over a (variety of) finite fields in order to make them tractable, as is standard in such situations.
\end{enumerate}

As briefly mentioned in the proceeding subsection, the use of characters in step $3$ here is crucial in rendering the above algorithm practically feasible. It enables the number of potential symmetries to be greatly reduced before it is necessary to specify the full CICY group action or use computationally intensive techniques such as Gr\"obner basis computations. Given its importance, we review some more details of this step, following \cite{Braun:2010vc}, in the next subsection.

\subsection{Fixed point freeness checks from character valued indices} \label{quickchecksec}

In this subsection, we present a detailed review of the character valued index methods used in step 3 of the algorithm discussed in the proceeding subsection. As we have seen, a CICY group action acts on the homogeneous coordinates of the ambient space of the Calabi-Yau manifold $X$ via the $\pi$ representation $(G,\overrightarrow{d},\pi_r,\gamma)$. This action defines a map from $X$ to itself, $\gamma(g): X\rightarrow X$, $\forall g\in G$. For each action $\gamma(g)$, there is an induced action on an invariant line bundle $\cal{L}$ with respect to this group action as well, $\gamma^{*}(g)$, $\gamma^{*}(g)\cal{L}\rightarrow \cal{L}$. Finally, this pullback map will induce an action on the cohomology $H^i(X,\cal{L})$ of the $G$ invariant line bundle $\cal{L}$ which, by a slight abuse of notation, we will denote by the same symbol. Given all of this, the following character valued generalization of the Euler characteristic of a line bundle ${\cal L}$ can be defined.
\begin{eqnarray}\label{inducedcharld}
\chi(g)(\mathcal{L})=\sum_i(-1)^i\text{Tr}_{H^i(\mathcal{L})}(\gamma^{*}(g))
\end{eqnarray}
This is a character as it gives a number for each group element $g\in G$. It is a generalization of the Euler characteristic because if, for example, $\gamma(g)$ acts trivially on $X$, then $\chi(g)(\mathcal{L})$  just reduces to that more familiar object.

The reason it is useful to define (\ref{inducedcharld}) is that, {\it if the symmetry action is fixed point free}, then this quantity obeys the following two useful properties\footnote{We have simplified the statement of these properties here slightly from the general case to versions that hold for the symmetries we study. For the general statement see \cite{Braun:2010vc}.}.
\begin{itemize}
\item If $\mathcal{L}$ is an invariant line bundle with respect to the symmetry, then $\chi(\mathcal{L})(g)=0 \;\; \forall \;g \neq 1$.
\item If $\mathcal{L}$ is in addition equivariant, then $\chi(\mathcal{L})(1)$ should be an integer multiple of the order of the group.
\end{itemize}

\vspace{0.1cm}

Crucially, the character valued index (\ref{inducedcharld}) can be computed purely in terms of manipulations of characters, given the characters associated to the CICY group action being studied. Thus by computing (\ref{inducedcharld}) for judicious choices of ${\cal L}$ one can rule out many possible symmetries in an efficient manner because they have fixed points. As we will discuss further in the next section, there are a series of tricks in terms of choosing which ${\cal L}$'s to consider that can greatly speed up computation. Obviously, one should consider line bundles for which less calculation resources are required to compute (\ref{inducedcharld}) first, as once a symmetry has been shown to have fixed points there is no need to proceed further. More subtle improvements can be obtained by recalling that, as we saw in former sections, a group action will separate the projective spaces factors of the ambient space into several orbits. It is possible to choose ${\cal L}$'s in such a manner that they only are affected by the action on a single orbit. In this manner orbits can be checked one-by-one and if an action on an orbit is found not to be compatible with fixed point freeness then one can rule out its inclusion in any Cartesian product to give a full CICY group action. Such methodology can show that large numbers of possible symmetries are not fixed point free with one computation of the character valued index.

Utilizing these techniques, it turns out that computing the character valued index (\ref{inducedcharld}) for various ${\cal L}$ can be used to rule out most possible symmetries which actually exhibit fixed points (a fact that can be gleaned by comparing the results of steps 3 and 4 of the algorithm in Section \ref{mralg}). Such computations are key to making a classification of the symmetries of interest tractable and thus, in the following subsections, we will describe in some detail, following \cite{Braun:2010vc}, how they are performed.

\subsubsection{Character valued cohomology on a single projective space}\label{secmroneP}

Let us start by describing how to compute the character valued cohomology $\mathfrak{h}^q(\mathbb{P}^j, {\cal L})$ of a line bundle ${\cal L}$ on $\mathbb{P}^j$ efficiently. We consider the case where we are given a group action on the homogeneous coordinates of $\mathbb{P}^j$ and wish to compute the induced representation on the $q$'th cohomology of ${\cal L}$, $H^q(\mathbb{P}^j, {\cal L})$. The quantity $\mathfrak{h}^q(\mathbb{P}^j, {\cal L})$ is then simply the vector given by the trace of this representation for each group element. This is a character, hence the name character valued cohomology, which encodes the information of the representation content of the line bundle cohomology being considered.

In principle, one could write down a polynomial (or more precisely Bott-Borel-Weil tensorial \cite{Hubsch:1992nu}) descriptions for $H^q(\mathbb{P}^j, {\cal L})$. Then it is possible to induce an action of the symmetry on these polynomial descriptions from the symmetry action on the $\mathbb{P}^j$ coordinates. Once these symmetry actions have been obtained, one could then directly compute the trace of the representation matrices. Given that our computations are going to become much more complex, however, we follow \cite{Braun:2010vc} in performing the computation directly in terms of the character $\chi$ of the action of the homogeneous coordinates of $\mathbb{P}^j$.

On $\mathbb{P}^j$ a line bundle cohomology $H^q(\mathbb{P}^j, {\cal L})$ can be non vanishing for $q=0$ or $q=j$. We will deal with each of these cases in turn. Global holomorphic sections of the  line bundle $ {\cal L}= {\cal O}(k)$ where $k \geq 0$ on $\mathbb{P}^j$ can be represented by the set of degree $k$ polynomials in the homogeneous coordinates of that projective space. The group action on the homogeneous coordinates will form some representation $r$ with character $\chi_r$. The action on these symmetric polynomials is then associated to the character $\chi_{\text{Sym}^k(r)}$ associated to the $k^\textnormal{th}$ symmetric power of $r$. 
\begin{eqnarray} \label{h0pj}
\mathfrak{h}^0 (\mathbb{P}^{j}, {\cal O}(k))=\chi_{\text{Sym}^k (r)}.
\end{eqnarray}
Useful formulae for computing $\chi_{\text{Sym}^k(r)}$ for different $k$ can be found in Appendix \ref{mrappendix}.

For the cohomology $H^j(\mathbb{P}^j, {\cal O}(-k))$ where $k \geq0$ the formula is a little more involved. Cohomologies of this type can be represented as completely antisymmetric epsilon symbols multiplied by polynomials of degree $k-j$ in the inverses of the homogeneous coordinates $1/x_i$ with $i=0,\ldots,j$ \cite{Hubsch:1992nu}. We then have the following.
\begin{eqnarray} \label{hn0pj}
\mathfrak{h}^j(\mathbb{P}^{j}, {\cal O}(-k))= \chi_\epsilon  \left( \chi_{\text{Sym}^{k-j-1}(r^{-1}) } \right)
\end{eqnarray}
Here $r^{-1}$ is the representation of the symmetry action on the inverse of the homogeneous coordinates $1/x_i$ and the number $\chi_\epsilon$ is given by the action on the completely antisymmetric epsilon symbol $\epsilon_{0\ldots j}$ induced from that on the homogeneous coordinates. 

\vspace{0.3cm}
\noindent
{\bf Example:}

\vspace{0.2cm}

To illustrate the above, let us consider the simple example of inducing an action of $\mathbb{Z}_2$ on $H^0(\mathbb{P}^1, {\cal O}(2))$. Let us take the group action on the homogeneous coordinates of $\mathbb{P}^1$ to be given by the following.
\begin{eqnarray} \label{eg1rep1}
\gamma (1) = \left( \begin{array}{cc} 1&0 \\0&1 \end{array} \right) \;\;,\;\; \gamma(g) = \left( \begin{array}{cc} 1&0 \\0&-1 \end{array} \right) \;\;\;\;
\end{eqnarray}
A basis for the global section of ${\cal O}(2)$ is given by $\left\{ x_0^2 ,x_0 x_1, x_1^2 \right\}$ and, acting on this basis, the induced group action can easily be seen to be of the following form.
\begin{eqnarray} \label{mr2rep}
\gamma_{H^0({\cal O}(2))}(1) = \left( \begin{array}{ccc} 1&0&0 \\0&1&0\\0&0&1 \end{array} \right) \;\;,\;\;\gamma_{H^0({\cal O}(2))}(g) = \left( \begin{array}{ccc} 1&0&0 \\0&-1&0\\0&0&1 \end{array} \right) \;\;\;\;
\end{eqnarray}

\vspace{0.1cm}

Let us reproduce this simple result using (\ref{h0pj}). The character associated to (\ref{eg1rep1}) is $\chi_r=(2,0)$. From (\ref{sym2}) in Appendix \ref{mrappendix}, we see that the formula for $\chi_{\text{Sym}^2(r)}$ is as follows.
\begin{eqnarray} \label{symeg1}
\chi_{\text{Sym}^2(r) } (g)= \frac{1}{2} \left( \chi_r(g)^2+\chi_r(g^2)\right)
\end{eqnarray}
In our case we have $\chi_r(g)^2=(4,0)$ and $\chi_r(g^2)= (2,2)$ (since $g^2=1$ for $\mathbb{Z}_2$). Thus, a simple computation using (\ref{h0pj}) and (\ref{symeg1}) returns that $\mathfrak{h}^0(\mathbb{P}^{j}, {\cal O}(2))=\chi_{\text{Sym}^2(r) } =(3,1)$ in this case. This result agrees precisely with the trace of the representation matrices given in (\ref{mr2rep}).
\subsubsection{Character valued indices on products of projective spaces}
Now let us go to the next layer of complexity of calculating the character valued index and consider line bundles on products of projective spaces. Here, one important difference from the single projective space case is that the group can not only act on the homogeneous coordinates but can also permute projective space factors with the same dimension. We will first describe how to compute the character of the induced representation on a single orbit of a group action. Then the generalization to allow for multiple distinct orbits will be straightforward by using Kunneth's formula. 

\vspace{0.2cm}

Consider a non-trivial orbit involving multiple copies of a complex projective space. As in previous subsections, we first consider the action $\gamma_1|_{G_1}$ of the stabilizer $G_1$ on the homogeneous coordinates of the first complex projective space factor. Then, instead of computing the induced action on the homogeneous coordinates of the entire orbit, as we did in Section \ref{homogcoordeg}, we wish to compute the induced action on a line bundle cohomology associated to the orbit. To do this, we induce the required result from character valued cohomologies associated to single projective spaces as computed in (\ref{h0pj}) and (\ref{hn0pj})

Consider an orbit $\prod_i \mathbb{P}^{n_i}$ of a projective space factor $\mathbb{P}^{n_1}$ under some $\pi$ representation with row action $\pi_r$. Define the embedding map 
\begin{eqnarray}
f_1: \mathbb{P}^{n_1} \to \prod_i \mathbb{P}^{n_i}\;.
\end{eqnarray}
It can be proven that the correct operation to compute the character of the induced representation of $G$ on a cohomology associated to the whole orbit, $\mathfrak{h}^{|G:G_1| q} \left( \prod_i \mathbb{P}^{n_i}, {\cal L} \right)$ from the row permutations $\pi_r$ and the character of the representation of the stabilizer $G_1$ acting on a cohomology associated to the first projective space $\mathfrak{h}^p(\mathbb{P}^{n_1}, f_1^*{\cal L})$ is given by the following \cite{Braun:2010vc}. 
\begin{eqnarray} \label{productoneorbit}
\mathfrak{h}^{|G:G_1| q} \left( \prod_i \mathbb{P}^{n_i}, {\cal L} \right)= \textnormal{GrInd}_{G_1}^G(\mathfrak{h}^q(\mathbb{P}^{n_1},f_1^*{\cal L}))
\end{eqnarray}
In this formula, the operation $\textnormal{GrInd}$ is equivalent to $\textnormal{SymInd}$ if $q$ is even and $\textnormal{AltInd}$ if $q$ is odd, symmetrized and antisymmetrized power versions of the induction operation respectively. This is simply because, thinking of elements of cohomology as equivalence classes of forms, even degree forms commute, leading to a symmetric product of representations of which we must take the character, whereas odd degree forms anti-commute, leading to the antisymmetric operation instead. Formulae for $\textnormal{SymInd}$ and $\textnormal{AltInd}$ can be found in Appendix \ref{mrappendix}. It should be noted that the degree of the character valued cohomology on the left of (\ref{productoneorbit}) is $|G:G_1| q$. The reason for this is that we are combining $q$ degree cohomologies for $|G:G_1|$ projective spaces in the orbit. Given (\ref{productoneorbit}), and the results from Section \ref{secmroneP}, we can compute character valued cohomology on a product of projective spaces forming an orbit of a group action.

\vspace{0.3cm}

In the above discussion we described how the operation $\text{GrInd}$ can be used to calculate character valued cohomology of a line bundle on a single orbit of projective spaces. More generally, the ambient space $A=\Pi_{k=1}^n\mathbb{P}^{d_k}$ of a CICY will be the product of several orbits. Given a $\pi$ representation $(G, \pi_r, \overrightarrow{d}, \gamma)$ acting on it, $A$ will be split into several orbits as follows:
\begin{eqnarray}
S_n=\{1\dots\}\cup\dots\cup\{\dots n\}=\bigcup_{G\{i\}\in S_n/G} G\{i\}.
\end{eqnarray}
To calculate the character valued cohomology of an invariant line bundle $\mathcal{L}$ on $A$, we will follow the following procedure \cite{Braun:2010vc}. First, for each orbit, we find a representative projective space $\mathbb{P}^{d_i}$, and we denote its embedding in $A$ by $i$.
\begin{eqnarray}
i: \mathbb{P}^{d_i}\longrightarrow A
\end{eqnarray}
Let us label the stabilizer of $\mathbb{P}^{d_i}$ by $G_i$. By following the procedure described above we can obtain the character valued cohomology associated to each orbit. A simple application of the Kunneth formula then reveals that we can obtain the character valued cohomology on the total ambient space $A$ by taking a simple product of those results.
\begin{eqnarray} \label{productorbit}
\mathfrak{h}^{q}(A,\mathcal{L})=\prod_{G\{i\}\in S_n/G}\text{GrInd}^G_{G_i}(\mathfrak{h}^{q_i}_i(\mathbb{P}^{d_i},i^{*}\mathcal{L})).
\end{eqnarray}
In this expression,
\begin{eqnarray}
q= \sum_{G\{i\}\in S_n/G} |G:G_i| q_i ,
\end{eqnarray}
with the cohomological degrees of the pieces of the product in (\ref{productorbit}) adding, as is usual in the Kunneth formula.

\vspace{0.2cm}

In what follows it will be useful to calculate, not just the character valued cohomology, but in addition the character valued index of invariant line bundles on $A$. This is a straightforward application of (\ref{productorbit}) to (\ref{inducedcharld}) which leads us immediately to the following result.
\begin{eqnarray}\label{chionp}
\chi(A,\mathcal{L})=\sum_{\vec{q}}(-1)^{\sum_{G\{i\}\in S_n/G} |G:G_i|q_i}\prod_{G\{i\}\in S_n/G}\text{GrInd}^G_{G_i}(\mathfrak{h}^{q_i}_i(\mathbb{P}^{d_i},i^{*}\mathcal{L}))
\end{eqnarray}
In the above formula $\vec{q}$ is a degree vector composed by all the $q_i$. 

\vspace{0.3cm}
\noindent
{\bf Example:}

\vspace{0.2cm}

Let us consider the example of ${\cal L}= {\cal O}(-3,-3)$ on $\mathbb{P}^1 \times \mathbb{P}^1$ with the symmetry action considered in (\ref{fullrepeg}). This $\mathbb{Z}_4$ symmetry action, which is associated with a $\pi_r$ where the generator of $\mathbb{Z}_4$ swaps the two $\mathbb{P}^1$ factors, is induced from the action (\ref{smallrep}).

Let us take as a basis for $H^2(\mathbb{P}^1 \times \mathbb{P}^1, {\cal O}(-3,-3))$ the set\footnote{We should really use Bott-Borel-Weil tensor descriptions of the cohomology here. However, the result is unchanged from that obtained from this rationome description and so we will stick to this form here for simplicity}
\begin{eqnarray} \label{egbasis}
\frac{1}{x_0 x_1 y_0 y_1} \times \left\{ \frac{1}{x_0 y_0} , \frac{1}{x_0 y_1} , \frac{1}{x_1 y_0} , \frac{1}{x_1 y_1} \right\}\;.
\end{eqnarray}
Here we have denoted the homogeneous coordinates of the first $\mathbb{P}^1$ by $x_0,x_1$ and those of the second $\mathbb{P}^1$ by $y_0, y_1$. Direct computation then allows us to derive the action induced from (\ref{fullrepeg}) on $H^2(\mathbb{P}^1 \times \mathbb{P}^1, {\cal O}(-3,-3))$. In terms of the basis (\ref{egbasis}) we find the following action.
\begin{eqnarray} \label{prodsecegaction}
\gamma_{H^2({\cal O}(-3,-3))}(1)= \left( \begin{array}{cccc} 1&0&0&0\\0&1&0&0 \\ 0&0&1&0\\ 0&0&0&1\end{array}\right) \;\;,\;\;\gamma_{H^2({\cal O}(-3,-3))}(g)= \left( \begin{array}{cccc} 1&0&0&0\\0&0&1&0 \\ 0&-1&0&0\\ 0&0&0&-1\end{array}\right) \\ \nonumber
\gamma_{H^2({\cal O}(-3,-3))}(g^2)= \left( \begin{array}{cccc} 1&0&0&0\\0&-1&0&0 \\ 0&0&-1&0\\ 0&0&0&1\end{array}\right) \;\;,\;\;\gamma_{H^2({\cal O}(-3,-3))}(g^3)= \left( \begin{array}{cccc} 1&0&0&0\\0&0&-1&0 \\ 0&1&0&0\\ 0&0&0&-1\end{array}\right) 
\end{eqnarray}

\vspace{0.1cm}

Let us now perform the computations involving characters outlined in this section and show that we get the result that one would expect from the above direct calculation. We consider the action of the stabilizer $\mathbb{Z}_2$ of the first $\mathbb{P}^1$ factor, as given in (\ref{smallrep}). First, we use (\ref{hn0pj}) to obtain the character valued cohomology associated to $H^1(\mathbb{P}^1 , {\cal O}(-3))$.
\begin{eqnarray}
\mathfrak{h}^1(\mathbb{P}^1, {\cal O}(-3))= \chi_\epsilon  \left( \chi_{\text{Sym}^{1}(r^{-1}) } \right) = (1,-1) \left(2,0 \right)=(2,0)
\end{eqnarray}
Once we have obtained $\mathfrak{h}^1(\mathbb{P}^1, {\cal O}(-3))$, we can use (\ref{productoneorbit}). In this example, we have $q$ from (\ref{productoneorbit}) taking a value of $1$, and thus $\textnormal{GrInd}$ is the same as $\textnormal{AltInd}$ in this case. In addition, $G=\mathbb{Z}_4$, $G_1=\mathbb{Z}_2$, $|G:G_1|=|\mathbb{Z}_4:\mathbb{Z}_2|=2$ and $f_1^*{\cal L}={\cal O}(-3)$. Therefore, we find the following from (\ref{productoneorbit}).
\begin{eqnarray} \label{thiscaseorb}
\mathfrak{h}^2(\mathbb{P}^1 \times \mathbb{P}^1 ,{\cal O}(-3,-3)) &=& \textnormal{AltInd}_{\mathbb{Z}_2}^{\mathbb{Z}_4} (\mathfrak{h}^1(\mathbb{P}^1,{\cal O}(-3))) \\ \nonumber
&=& \textnormal{AltInd}_{\mathbb{Z}_2}^{\mathbb{Z}_4} ((2,0))
\end{eqnarray}
Using (\ref{indwith2}) together with some intuitive notation introduced in Appendix \ref{mrappendix}, we find that
\begin{eqnarray} \label{theoneweneed}
 \textnormal{AltInd}_{\mathbb{Z}_2}^{\mathbb{Z}_4} ((2,0)) = \text{Alt}^2 (\text{Ind}_{\mathbb{Z}_2}^{\mathbb{Z}_4}((2,0)) )- \text{Ind}_{\mathbb{Z}_2}^{\mathbb{Z}_4}(\text{Alt}^2 ((2,0)))\;.
 \end{eqnarray}
We have already computed $\text{Ind}_{\mathbb{Z}_2}^{\mathbb{Z}_4}((2,0))=(4,0,0,0)$ in the discussion around equations (\ref{char1}) to (\ref{badger3}). We then compute, using (\ref{alt2}) in Appendix \ref{mrappendix}:
\begin{eqnarray} \label{mraltagain}
\chi_{\text{Alt}^2(r)} (g) = \frac{1}{2} \left[ \chi_r(g)^2 - \chi_r(g^2) \right] \;,
\end{eqnarray}
to find the following. 
\begin{eqnarray} \label{term1}
\text{Alt}^2 (\text{Ind}_{\mathbb{Z}_2}^{\mathbb{Z}_4}((2,0))) = \text{Alt}^2 ((4,0,0,0)) = \frac{1}{2} \left[ (16,0,0,0)-(4,0,4,0)\right] =(6,0,-2,0)
\end{eqnarray}
Similarly, given (\ref{mraltagain}) we find
\begin{eqnarray}
\text{Alt}^2 ((2,0))=\frac{1}{2} \left[ (4,0)-(2,2) \right] = (1,-1)
\end{eqnarray}
Finally, pursuing a computation to work out the induced character as we did in the discussion around (\ref{char1}) to (\ref{badger3}) we arrive at the following.
\begin{eqnarray} \label{term2}
\text{Ind}_{\mathbb{Z}_2}^{\mathbb{Z}_4}( \text{Alt}^2 ((2,0))) = (2,0,-2,0)
\end{eqnarray}

Putting the above discussion together and combining our results in (\ref{term2}), (\ref{term1}), (\ref{theoneweneed}) and (\ref{thiscaseorb}) we obtain the following.
\begin{eqnarray}
\mathfrak{h}^2(\mathbb{P}^1 \times \mathbb{P}^1 ,{\cal O}(-3,-3))= \textnormal{AltInd}_{\mathbb{Z}_2}^{\mathbb{Z}_4} ((2,0)) =(6,0,-2,0) - (2,0,-2,0)=(4,0,0,0)
 \end{eqnarray}
Comparing this result to (\ref{prodsecegaction}) we see that it correctly reproduces the traces of the representation matrices.

\subsubsection{Character valued indices on CICYs}

In the above we have described how to obtain character valued indices of line bundles on the ambient space $A$ of our Calabi-Yau manifolds. The final step is to use these results to obtain expressions for the character valued indices of line bundles on the threefold $X$ itself. Line bundles on the ambient space and Calabi-Yau manifold are related to one another by the long exact Koszul sequence.
\begin{eqnarray} \label{koszul}
0\rightarrow\mathcal{O}(-\sum D_j)\otimes V\rightarrow\dots\rightarrow(\bigoplus_{j<k}\mathcal{O}(-D_j-D_k))\otimes V\rightarrow \mathcal{O}(-D_i)\otimes V\rightarrow V|_X\rightarrow 0,
\end{eqnarray}
Here, all of the objects are vector bundles on $A$ except $V|_X$ which is the sheaf on $A$ which corresponds to the associated vector bundle on $X$. The $D_i$ are the divisor classes of the $i'th$ defining relations of the CICY. By the properties of long exact sequences, if the defining relations of the Calabi-Yau manifold, and thus the maps in (\ref{koszul}), were to be $G$-invariant then one simply obtain the following.
 \begin{eqnarray} \label{chisum}
\chi(X, \mathcal{L}|_X)=\sum_{1<j_1<j_2<\dots<j_p}(-1)^p\chi( \mathcal{L}\otimes\mathcal{O}(-D_{j_1}-D_{j_2}-\dots-D_{j_p}))|_{A}.
\end{eqnarray}
However, in general the defining relations, and thus maps, involved do transform non-trivially, and thus more care is needed in computing the character valued index on $X$.

Following \cite{Braun:2010vc}, we define some notation to make the formulae that follow more compact. First we denote ${\cal O}(-D_{j_1}-\ldots - D_{j_p})= {\cal O}_{j_1 \wedge \ldots \wedge j_p}$. We then define the standard basis of anti-symmetrized indices as follows.
\begin{eqnarray} 
\Lambda_m = \left\{ j_1 \wedge \ldots  \wedge j_p | 0 \leq p \leq m, 1 \leq j_1 < \ldots < j_p\leq m\right\}
\end{eqnarray}
Unfortunately, a line bundle ${\cal O}_{j_1 \wedge \ldots \wedge j_p}$ will not in general be invariant under our symmetry action, thus we must consider index sets which would group the associated bundles into orbits of $G$. We thus define,
\begin{eqnarray}
\Lambda_m /G= \left\{ [\wedge \vec{j}_{(1)}],[\wedge \vec{j}_{(2)}],\ldots\right\} = \left\{j_1 \wedge \ldots \wedge j_p | 0 \leq p\leq m \right\}/\langle \pm, G \rangle \;.
\end{eqnarray} 
Here $ [\wedge \vec{j}_{(i)}]$ denotes a class of $\wedge {\vec{j}}$ that form an orbit under $G$ and in the second equality, the relevant classes have been expressed in terms of representatives.

\vspace{0.1cm}

We denote by
\begin{eqnarray} \label{andjelly}
p_{\wedge \vec{j}} = p_{j_1} \wedge \ldots \wedge p_{j_p}
\end{eqnarray}
the exterior powers of the group characters associated to defining polynomials $j_1$ to $j_p$. It turns out that this anti-symmetric object plays a key role in the results to follow because of the way that negative signs that appear in the map structure of the Koszul sequence \cite{Braun:2010vc}.

\vspace{0.1cm}

We need to define some stabilizers of the symmetry group $G$. First we have the stabilizer of a particular multi index $\wedge \vec{j}$.
\begin{eqnarray}
G_{\wedge \vec{j}} = \text{stab}_{\wedge \vec{j}} (G) = \left\{  g\in G| \pi(g)(\wedge \vec{j}) = \pm \wedge \vec{j}\right\}
\end{eqnarray}
Second, we need the stabilizer of an individual defining relation.
\begin{eqnarray}
G_{ {j}} = \text{stab}_{ {j}} (G) = \left\{  g\in G| \pi(g)( {j}) = \pm  {j}\right\}
\end{eqnarray}

\vspace{0.1cm}

With these preliminaries completed we can now describe how (\ref{chisum}) is modified by the $G$ dependence of the maps in the Koszul sequence (\ref{koszul}). We can compute $p_{\wedge \vec{j}}$ defined in (\ref{andjelly}) via the following formula.
\begin{eqnarray} \label{orbits1p}
p_{\wedge\overrightarrow{j}}=\prod_{j\in \overrightarrow{j}/G}\text{AltInd}^{G_{\wedge\overrightarrow{j}}}_{G_j\cap G_{\wedge\overrightarrow{j}}}(\text{Res}^{G_j}_{G_j\cap G_{\wedge\overrightarrow{j}}}(p_j)).
\end{eqnarray}
This computes the relevant $G_{\wedge \vec{j}}$ character from $G_j$ characters associated to single defining polynomials. Given these $G_{\wedge \vec{j}}$ characters we can then compute the $G$-character valued index of the line bundle ${\cal L}|_X$ on $X$ as follows \footnote{In fact, the final formula given in \cite{Braun:2010vc} is somewhat more complicated than the one reproduced here. For the types of symmetry we are studying in this paper, this simpler expression turns out to be sufficient.}.
\begin{eqnarray} \label{chisumproper}
\chi({\cal L}|_X) = \sum_{\wedge \vec{j} \in \Lambda_m/G} (-1)^{|\wedge \vec{j}|} \text{Ind}^G_{G_{\wedge \vec{j}}} \left(  p_{\wedge \vec{j}} \chi ({\cal O}_{\wedge \vec{j}} \otimes {\cal L})\right)
\end{eqnarray}
This formula generalizes (\ref{chisum}) in two ways. First it performs the sum not over line bundles but over orbits of line bundles appearing in the Koszul sequence. The induction that appears in (\ref{chisumproper}) induces the action on an entire orbit from the action of the stabilizer group on a single element in the orbit. Second, (\ref{chisumproper}) takes into account the non-trivial character structure due to the polynomials themselves, though $p_{\wedge \vec{j}}$ as given in (\ref{orbits1p}). Note that both $p_{\wedge \vec{j}}$ and $\chi ({\cal O}_{\wedge \vec{j}}\otimes {\cal L})$ are $G_{\wedge \vec{j}}$ characters so this formula induces a $G$ character from an $G_{\wedge \vec{j}}$ one correctly as indicated by the use of $\text{Ind}^G_{G_{\wedge \vec{j}}}$.

\vspace{0.3cm}

With formula (\ref{chisumproper}) we have a closed form expression for computing the character valued index of line bundles on CICYs. This is all that is required to perform step 3 of the algorithm given in Section \ref{mralg}. 

\section{Results} \label{resultssec}

In this section we will describe the results of our scan for cyclic symmetries on the favorable CICY list. In doing so we will emphasize comparisons to the symmetries found in \cite{Braun:2010vc} which were manifest from the description of the same manifolds found in the standard CICY list. Our code reproduced identical results to those found in \cite{Braun:2010vc} for the cyclic symmetries when applied to the original CICY list, providing a highly non-trivial check that the code is working as intended before applying it to new, favorable, configuration matrices. In obtaining the results in this paper we have made use of the CICY Package \cite{CICYPackage}, GAP \cite{GAP4} and Singular \cite{singular}.

This section is split into four subsections. In the first, we describe how to access our full results from an associated online database and provide some overview statistics of the types of symmetries, and so non-simply connected Calabi-Yau threefolds, we have found. In subsequent subsections we give examples illustrating how the symmetries which are manifest from ambient space actions in the favorable CICY list can both vary from and overlap with those manifest from the original CICY list.

\subsection{Statistics and database of results} \label{statres}

\begin{figure}[t]
\centering
\includegraphics[width=1\textwidth]{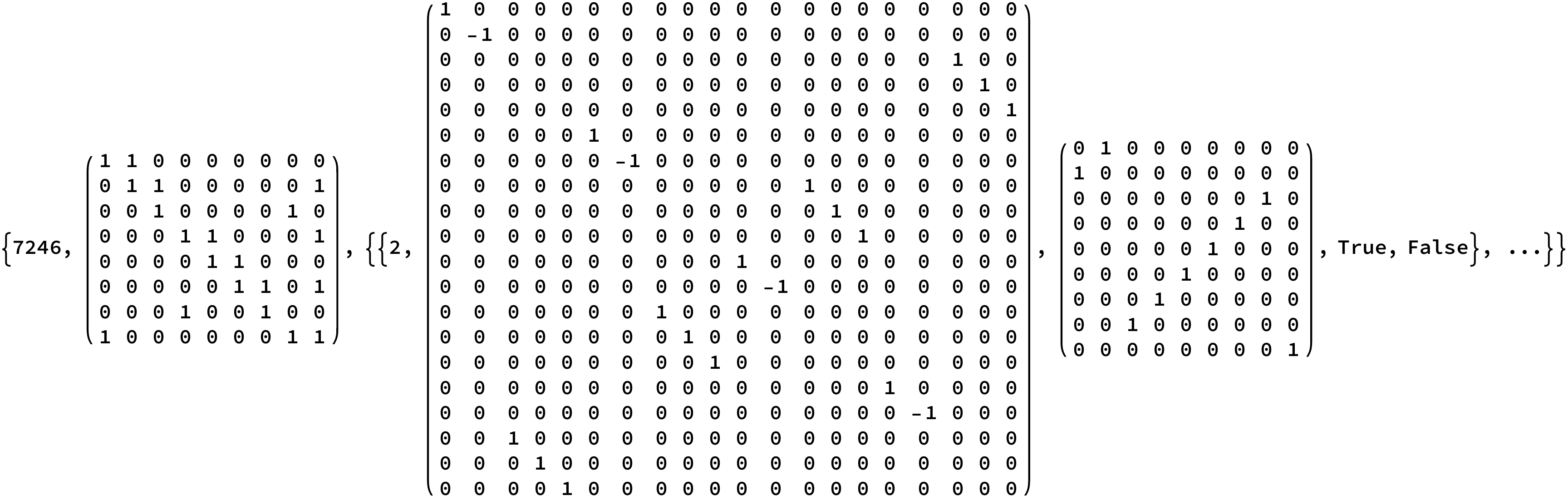}
\caption{ \emph{ A simple example illustrating the format described in the text. This example describes the symmetries associated to CICY 7246, the favorable configuration matrix for which is the second entry in this list. The third entry in the list is a list of symmetries of which only the first is shown explicitly. This symmetry is associated to an order $2$ cyclic group with the action on the homogeneous coordinates being given by the first matrix and the action on the defining polynomials being given by the second. This particular example has been proven to lead to a smooth set of invariant defining relations but it is not know whether it corresponds to a new symmetry (although it is certainly a new representation of one).}}
\label{formateg}
\end{figure}

The symmetries we have classified can be found at \href{http://www1.phys.vt.edu/~grayphys/CyclicSyms.html}{this link}, in the form of a machine readable list. Each entry in the list is of the following form.
\begin{eqnarray} \nonumber
\text{\{Num,Config,\{\{$n_1$,$\text{CoordAct}_1$,$\text{PolyAct}_1$, $\text{Smooth}_1$,$\text{New}_1$\},\{$n_2$,$\text{CoordAct}_2$,$\text{PolyAct}_2$,$\text{Smooth}_2$,$\text{New}_2$\},$\ldots$\}\}}
\end{eqnarray}
Here $\text{Num}$ is the number of a CICY in the original CICY list, and $\text{Config}$ is the favorable configuration matrix for that manifold as provided in \cite{Anderson:2017aux}. The third entry of the list is a collection of lists, one for each symmetry manifest from the favorable configuration. The integer $n_\#$ indicates that the cyclic group $\mathbb{Z}_{n_\#}$ is being manifested. The entries $\text{CoordAct}_{\#}$ and $\text{PolyAct}_{\#}$ are the actions of the homogeneous coordinates of the ambient space and the defining polynomials of the favorable CICY respectively. Finally the entries $\text{Smooth}_{\#}$ and $\text{new}_{\#}$ will either by $\text{True}$ or $\text{False}$. As we will discuss shortly, in some cases the computational power we have had available has proven insufficient to decide whether the manifold where the complex structure has been tuned in order to respect the symmetry is smooth or not. If the manifold has proven to be smooth then $\text{Smooth}_{\#}$ will be $\text{True}$ otherwise it will be $\text{False}$. Cases that have been proven to be singular have of course be discarded. The entry $\text{new}_{\#}$ will be $\text{True}$ if it has been proven that the symmetry in question is not equivalent to one in the results of \cite{Braun:2010vc}, otherwise it will be $\text{False}$. This has simply been decided by determining whether a symmetry of the correct order on the manifold in question appeared in \cite{Braun:2010vc} or not. This is a very crude, sufficient but not necessary, condition for symmetries to be previously unknown, which may be improved upon in future work as described below. An example of this format can be found in Figure \ref{formateg}.

\vspace{0.2cm}

There are 2946 CICYs for which a new, favorable, configuration matrix is provided in the list of \cite{Anderson:2017aux}. Of these, 1393 manifolds have at least one symmetry that would be in principle allowed by index considerations (step 1, in the algorithm in Section \ref{mralg}).
Scanning over those 1393 manifolds, we found, using the character valued index methods of step 3 in the algorithm, that there are 199 possible fixed point free symmetries on 99 different manifolds. Of these almost all, 175 out of 199 on the same 99 manifolds, are really fixed point free when the necessary and sufficient check of algorithm step 4 was performed. These numbers show how efficient the character valued index method developed in \cite{Braun:2010vc} is at deciding this issue. After removing some redundancies, where symmetries can be shown to be the same by trivial relabeling of ambient space factors and defining relations, the total number of symmetries is reduced to 129. Finally a smoothness check was carried out as per step 5 of the algorithm. Some of these computations are still continuing and it is not known if they will finish with currently available computing resources (the favorable list involves bigger configuration matrices than the original CICY list, making such computations much more intensive). Thus far we have shown that approximately half of these symmetries do lead to smooth invariant defining relations. For the remaining examples, results will be added to the online data base when and if they become available. Of the different possible types of freely acting cyclic symmetries we find 101 $\mathbb{Z}_2$,  14 $\mathbb{Z}_3$ and 14 $\mathbb{Z}_4$ symmetries.

It is interesting to compare the symmetries found for these 2946 favorable configurations with those descending from the ambient space of the description of the same manifolds in the original CICY list. There, in \cite{Braun:2010vc},  506 cyclic symmetries were found over 97 manifolds. These included, 432 $\mathbb{Z}_2$, 28 $\mathbb{Z}_3$, 14 $\mathbb{Z}_4$, 27 $\mathbb{Z}_6$ and 5 $\mathbb{Z}_{12}$ symmetries. Clearly there are differences between this symmetry content and what we have found for the favorable configurations. Many known symmetries do not descend from linear actions on the ambient space of the favorable matrices as there are 377 more symmetries found in the original list than in our scan. In particular, we find no $\mathbb{Z}_6$ or $\mathbb{Z}_{12}$ symmetries at all. By contrast there are some symmetries that descend from the ambient space of the favorable configuration matrices that are not present in the original list. In fact, we find 33 symmetries where a symmetry of the same order was not previously known on the manifold in question. These symmetries are definitely new. To know how much overlap there is between the remaining symmetries in the two lists would be a much longer question to answer. We give an example of how to address this issue in Section \ref{cosym}. 

\vspace{0.1cm}
From the above statistics we see that we have discovered, at the very least, 33 seemingly new non-simply connected Calabi-Yau manifolds, and possibly many more. One way to  gain some information as to which of the other geometries correspond to truly new topological types of Calabi-Yau threefold would be to compute invariants of these quotients, such as their Hodge numbers. The technology developed in \cite{Candelas:2008wb,Candelas:2010ve,Candelas:2015amz,Candelas:2016fdy,Constantin:2016xlj} could be used to do this, although such computations are beyond the scope of this paper. Instead, in the rest of this section we will examine in more detail a few examples which illustrate the relationship between the symmetries descending from linear actions on the ambient spaces of the favorable configurations and those descending from the non-favorable matrices of the original CICY list.

\subsection{Ruling out non-permuting cyclic symmetries for the new favorable CICYs} \label{labelforconc}

In this subsection we wish to illustrate two points. First, by calculating the character valued index for carefully chosen line bundles, one can quickly rule out the existence of large classes of possible symmetries. Second, symmetries descending from the ambient space of a CICY configuration matrix in the original list will not necessarily be manifested as a similar such an action in a given favorable configuration.

\vspace{0.2cm}

Let us start by discussing a particular property of the new favorable CICY matrices introduced in \cite{Anderson:2017aux}. They admit no cyclic symmetries which do not permute the rows or columns of the configuration matrix. To show this we need to consider the symmetries $\mathbb{Z}_{2}$, $\mathbb{Z}_3$, $\mathbb{Z}_{5}$, $\mathbb{Z}_{7}$ and $\mathbb{Z}_{13}$. By index considerations from step 1 of the algorithm in Section \ref{mralg}, these are the only cyclic groups of prime order that could appear. The possible non-prime cyclic groups whose order is compatible with index considerations all have subgroups in the above list. Therefore if the above groups can't be realized as fixed point free symmetries with smooth invariant defining polynomials then those larger groups certainly can not either. Thus if we can rule out these prime possibilities then our discussion will be complete. A character valued index computation for carefully chosen line bundles can rule out these cases with very little computational effort.

To illustrate the method, we will consider the favorable description of CICY number $7300$:
 \begin{eqnarray}
X= \left[ \begin{array}{c|c c c c c c c} 
\mathbb{P}^2 & 1 & 1 &1 & 0 & 0 & 0 & 0  \\ 
\mathbb{P}^1 & 0 & 0 & 1& 0 & 0 & 0 & 1 \\
 \mathbb{P}^1 & 0 &0 & 0& 0 & 1 & 0 & 1\\ 
 \mathbb{P}^2 & 0 &0 & 0& 1 & 1 & 1 & 0\\
 \mathbb{P}^1 & 0 &1 & 0& 0 & 0 & 0 & 1\\
 \mathbb{P}^1 & 1 &0 & 0& 0 & 0 & 0 & 1\\
 \mathbb{P}^1 & 0 &0 & 0& 1 & 0 & 0 & 1\\
 \mathbb{P}^1 & 0 &0 & 0& 0 & 0 & 1 & 1\\
 \end{array}\right] \; . \label{conf7300}
\end{eqnarray}
As we can see from this configuration matrix, this manifold has $7$ defining polynomials. For some line bundle $\mathcal{L}^{\prime}$ on $X$, there will be at most $128$ line bundles in the Koszul sequence whose character valued index would need to be calculated in order to obtain $\chi(\mathcal{L}^{\prime}|_{X})$ from (\ref{chisumproper}). One way to simplify this calculation is to choose a special line bundle $\mathcal{L}$, for which most of the line bundles in the associated Koszul sequence have zero cohomology. One candidate is the following.
 \begin{eqnarray} \label{line7300}
 \mathcal{L}=\mathcal{O}(1, 1, 0, 0, 0, 0, 0, 0)
\end{eqnarray}
For this choice of ${\cal L}$ it can trivially be shown, using the Bott-Borel-Weil theorem \cite{Hubsch:1992nu}, that the only ambient space line bundles appearing in the Koszul sequence that do not have vanishing cohomology are ${\cal O}$ and $\mathcal{L}$, with all others falling into the so called `Bott gap'. Further more, because of the form of (\ref{line7300}), we need only specify the symmetry action on the coordinates of the first two ambient space factors in order to compute $\chi({\cal L}|_X)$.

Using this let us examine the possibility of non-permuting $\mathbb{Z}_2$ symmetries. By using the two simplifications above, one can very quickly, by hand, show that there is no choice of action on the leading $\mathbb{P}^2 \times \mathbb{P}^1$ factor of the ambient space of (\ref{conf7300}) that will lead to $\chi({\cal L}|_X)$ being compatible with the existence of such as symmetry, according to the discussion of Section \ref{quickchecksec}.

\vspace{0.1cm}

One might think that the above simplifying choice of ${\cal L}$ is a special feature of the configuration matrix (\ref{conf7300}). However, a direct scan over all of the new favorable CICY configuration matrices of \cite{Anderson:2017aux} which could possibly have an order 2 symmetry shows that each of them exhibits a line bundle that satisfies the two following constraints.
\begin{itemize}
\item First, $\mathcal{L}$ has most of its entries $0$, except for a $1$ entry on a single $\mathbb{P}^1$ and a $1$ entry on a single $\mathbb{P}^2$.
\item Second, the Koszul sequence of $\mathcal{L}$ only contains two ambient space line bundles with non-vanishing cohomology, ${\cal O}$ and $\mathcal{L}$.
\end{itemize}
Furthermore, in each case the line bundle of interest is simply one of those appearing in the normal bundle. Thus one can quickly repeat the same computation in all of those cases to find that there can be no $\mathbb{Z}_2$ symmetries descending from the ambient space of these favorable configurations without row or column swapping actions. In fact, our exhaustive scan shows that there is no $\mathbb{Z}_2$ symmetry descending from the ambient space of (\ref{conf7300}) at all, with or without permutation actions. This same Calabi-Yau threefold {\it does} exhibit a $\mathbb{Z}_2$ symmetry which can be found in the classification \cite{Braun:2010vc} based upon the original CICY list. It is simply the case that this symmetry does not descend from the ambient space in a simple manner in the new configuration.

\vspace{0.1cm}

The reason this is happening could well be due to the particular choice of splittings that are used in generating favorable configurations in \cite{Anderson:2017aux}. For example, in (\ref{conf7300}) two $\mathbb{P}^2$ factors have been introduced relative to the original configuration matrix.
\begin{eqnarray}
X_{\text{non-favorable}}=\left[ \begin{array}{c|ccc} \mathbb{P}^1 & 1&0&1 \\\mathbb{P}^1 & 0&1&1 \\\mathbb{P}^1 & 0&1&1 \\\mathbb{P}^1 & 0&1&1 \\ \mathbb{P}^1 & 1&0&1 \\ \mathbb{P}^1 & 1&0&1  \end{array} \right]
\end{eqnarray}
There are natural actions of $\mathbb{Z}_2$ on $\mathbb{P}^1$ with minimal dimensionality fixed point loci, by virtue of the fact that every coordinate can transform differently under a toric action. The same is not true for a $\mathbb{P}^2$ ambient factor. Two of the coordinates of $\mathbb{P}^2$ must transform in the same manner under $\mathbb{Z}_2$ and this leads to a larger fixed point set, that is then more likely to intersect the Calabi-Yau manifold. 

This discussion leads us back to the comment that was made in Section \ref{favcicysec} that the favorable configurations found in \cite{Anderson:2017aux} are not unique. One could imagine attempting to generate different favorable configurations if one was interested in a specific symmetry that was either known to exist from a non-favorable description or suspected to exist from index and other considerations. For example, if one was interested in $\mathbb{Z}_2$ symmetries, one might try to generate such a configuration utilizing only $\mathbb{P}^1$ splits.

\vspace{0.3cm}

For $\mathbb{Z}_3$ symmetries there is an exactly analogous story. Ruling out  $\mathbb{Z}_{5}$, $\mathbb{Z}_{7}$ and $\mathbb{Z}_{13}$ symmetries which do not permute the equations or ambient space factors is slightly different however. One can quickly show that in those cases a line bundle satisfying the two criteria described above can not always be found. Instead, in these cases, we search for a line bundle with the following two properties. 
\begin{itemize}
\item First, $\mathcal{L}$ has most of its entries $0$, except having a $1$ entry on a single $\mathbb{P}^1$, a $1$ entry on a single $\mathbb{P}^2$.
\item Second, all of the ambient space line bundles in the Koszul sequence must have vanishing cohomology except for $\mathcal{L}$ itself.
\end{itemize}
With these conditions being imposed, ${\cal L}$ will never be a line bundle in the normal bundle. Nevertheless, a line bundle of this type can easily be found for every relevant manifold.

From the second of these new conditions, we can see that the character valued index of $\mathcal{L}_{X}$ is totally decided by that of $\mathcal{L}_{A}$, along with any data associated to the maps in the Koszul sequence. Then because of the first condition, the calculation of $\chi(\mathcal{L}_X)$ boils down to a simple calculation involving $\chi(\mathcal{O}(1,1)_{\mathbb{P}^1\times\mathbb{P}^2})$. Using the methods outlined in Section \ref{bigscansec} it is then easy to show that for a line bundle satisfying the two conditions of the previous paragraph, it is not possible to have a zero character valued index for a non-trivial group element for groups $\mathbb{Z}_{5}$, $\mathbb{Z}_{7}$ or $\mathbb{Z}_{13}$. Thus no such symmetries exist which are fixed point free. This completes the proof that there are no cyclic symmetries that do not permute rows or columns of the configuration matrix on the new favorable CICY configurations of \cite{Anderson:2017aux}.

\subsection{An example of a symmetry appearing in both favorable and non-favorable configurations}\label{cosym}

Despite the restriction on cyclic actions discussed in the previous sub-section there certainly are symmetries appearing in the results of \cite{Braun:2010vc} that are reproduced non-trivially in our classification based on the favorable CICY list. In this section we will give an example of a symmetry that can be realized in terms of linear actions on the ambient spaces of both the standard and favorable configuration matrices in this manner. The CICY we are going to examine is number $7800$ in the standard list with configuration matrix:
 \begin{eqnarray} \label{conf7800}
X_{7800}= \left[ \begin{array}{c|c c} 
\mathbb{P}^1 & 1 & 1  \\ 
\mathbb{P}^1 & 1 & 1  \\
 \mathbb{P}^1 & 1 &1 \\ 
 \mathbb{P}^2 & 0 &3 \\
 \end{array}\right] \; .
\end{eqnarray}
The favorable cousin of the above configuration matrix is as follows.
 \begin{eqnarray} \label{conf7800f}
X_{7800 \text{\;favorable}}= \left[ \begin{array}{c|c c c c} 
\mathbb{P}^2 & 1 & 1 & 1 & 0  \\ 
\mathbb{P}^1 & 0 & 0 & 1 & 1  \\ 
\mathbb{P}^1 & 0 & 1 & 0 & 1 \\
 \mathbb{P}^1& 1 & 0 & 0 & 1 \\ 
 \mathbb{P}^2 & 0 & 0 & 0 & 3 \\
 \end{array}\right] 
\end{eqnarray}
The favorable matrix above was obtained in \cite{Anderson:2017aux} by using a $\mathbb{P}^2$ split of the first column of (\ref{conf7800}). In order to compare a symmetry appearing in both of these descriptions of this manifold, we label the homogeneous coordinates of the first $\mathbb{P}^2$ in (\ref{conf7800f}) as $(x_{0,1}, x_{0,2}, x_{0,3})$. For other projective spaces in both (\ref{conf7800}) and (\ref{conf7800f}) we use the same notation since they are identical in the two cases. We label their homogeneous coordinates as $(x_{i,1},x_{i,2}, \ldots)$ where $i=1,\ldots 4$.

In \cite{Braun:2010vc} a $\mathbb{Z}_3$ symmetry descending from the ambient space of $X_{7800}$  was found which acts on the homogeneous coordinates in the following manner.
 \begin{eqnarray} \label{conf7800a}
\gamma_1=\left(
\begin{array}{ccccccccc}
 0 & 0 & 1 & 0 & 0 & 0 & 0 & 0 & 0 \\
  0 & 0 & 0 & 1 & 0 & 0 & 0 & 0 & 0 \\
  0 & 0 & 0 & 0 & 1 & 0 & 0 & 0 & 0 \\
  0 & 0 & 0 & 0 & 0 & 1 & 0 & 0 & 0 \\
  1 & 0 & 0 & 0 & 0 & 0 & 0 & 0 & 0 \\
 0 & 1 & 0 & 0 & 0 & 0 & 0 & 0 & 0 \\
  0 & 0 & 0 & 0 & 0 & 0 & 1 & 0 & 0 \\
 0 & 0 & 0 & 0 & 0 & 0 & 0 & e^{\frac{2 i \pi }{3}} & 0 \\
  0 & 0 & 0 & 0 & 0 & 0 & 0 & 0 & e^{-\frac{2 i \pi }{3}} \\
\end{array}
\right)
\end{eqnarray}
This $\mathbb{Z}_3$ symmetry acts on the defining polynomials trivially. For $X_{7800 \text{\;favorable}}$, we find a $\mathbb{Z}_3$ symmetry which acts on the homogeneous coordinates as follows.
 \begin{eqnarray} \label{conf7800b}
 \gamma_2=\left(
\begin{array}{cccccccccccc}
 1 & 0 & 0 & 0 & 0 & 0 & 0 & 0 & 0 & 0 & 0 & 0 \\
 0 & e^{\frac{2 i \pi }{3}} & 0 & 0 & 0 & 0 & 0 & 0 & 0 & 0 & 0 & 0 \\
 0 & 0 & e^{-\frac{2 i \pi }{3}} & 0 & 0 & 0 & 0 & 0 & 0 & 0 & 0 & 0 \\
 0 & 0 & 0 & 0 & 0 & 1 & 0 & 0 & 0 & 0 & 0 & 0 \\
 0 & 0 & 0 & 0 & 0 & 0 & 1 & 0 & 0 & 0 & 0 & 0 \\
 0 & 0 & 0 & 0 & 0 & 0 & 0 & 1 & 0 & 0 & 0 & 0 \\
 0 & 0 & 0 & 0 & 0 & 0 & 0 & 0 & 1 & 0 & 0 & 0 \\
 0 & 0 & 0 & 1 & 0 & 0 & 0 & 0 & 0 & 0 & 0 & 0 \\
 0 & 0 & 0 & 0 & 1 & 0 & 0 & 0 & 0 & 0 & 0 & 0 \\
 0 & 0 & 0 & 0 & 0 & 0 & 0 & 0 & 0 & 1 & 0 & 0 \\
 0 & 0 & 0 & 0 & 0 & 0 & 0 & 0 & 0 & 0 & e^{\frac{2 i \pi }{3}} & 0 \\
 0 & 0 & 0 & 0 & 0 & 0 & 0 & 0 & 0 & 0 & 0 & e^{-\frac{2 i \pi }{3}} \\
\end{array}
\right)
\end{eqnarray}
We find that this symmetry also acts on the defining polynomials non-trivially:
 \begin{eqnarray} \label{conf7800c}
\gamma_{2p}= \left(
\begin{array}{cccc}
 0 & 0 & 1 & 0 \\
 1 & 0 & 0 & 0 \\
 0 & 1 & 0 & 0 \\
 0 & 0 & 0 & 1 \\
\end{array}
\right).
\end{eqnarray}

If we compare these two coordinate actions, (\ref{conf7800a}) and (\ref{conf7800b}), we find that, apart from the coordinate action of the first $\mathbb{P}^2$ of (\ref{conf7800f}), they are identical. Considering the invariant defining polynomials in the two cases, we then see that the last defining polynomials of both (\ref{conf7800}) and (\ref{conf7800f}) only depend on coordinates which will be acted on identically by (\ref{conf7800a}) and (\ref{conf7800b}), and are thus the same. For the first three defining relations in (\ref{conf7800f}) we observe that all three of them depend linearly on the coordinates of the first $\mathbb{P}^2$ structure. Therefore they are solved iff the determinental equation associated to these relations holds (the standard structure for an ineffective splitting). This determinental variety is identical to the first defining relation of (\ref{conf7800}), tuned to be invariant under the symmetry (\ref{conf7800a}). 

\vspace{0.1cm}

Thus we see, as would be expected from the matching of the coordinate actions (\ref{conf7800a}) and (\ref{conf7800b}), these two symmetries are the same. They are simply expressed in terms of two different descriptions of one Calabi-Yau threefold.

\subsection{An example of a new symmetry}\label{newsym}

In addition to symmetries that descend only from linear actions on the ambient spaces given in the original CICY list, or cases which descend from both the original and favorable descriptions of the manifolds, there are new symmetries which descend from the ambient spaces of the favorable configurations. These are symmetries of these threefolds that are, as far as we are aware, new to this work, and thus so are the associated non-simply connected quotient Calabi-Yau manifolds.

One example of a new symmetry is found on CICY $6925$. In the standard CICY list this Calabi-Yau manifold is described by the following configuration matrix.
 \begin{eqnarray} \label{conf6925}
X_{6925}= \left[
\begin{array}{c|ccc}
\mathbb{P}^1 & 1 & 1 & 0 \\
\mathbb{P}^1 & 1 & 0 & 1 \\
\mathbb{P}^1 & 2 & 0 & 0 \\
\mathbb{P}^1 & 2 & 0 & 0 \\
 \mathbb{P}^2 &0 & 1 & 2 \\
\end{array}
\right]
\end{eqnarray}
In \cite{Braun:2010vc}, it was shown that this manifold does not exhibit any freely acting discrete symmetries that descend from linear actions on this ambient space.

The favorable description of the same manifold, as given in \cite{Anderson:2017aux}, is as follows.
 \begin{eqnarray} \label{conf6925f}
X_{6925 \text{\;favorable}}=
\left[
\begin{array}{c|ccccccc}
\mathbb{P}^1 & 1 & 1 & 0 & 0 & 0 & 0 & 0 \\
\mathbb{P}^2 & 0 & 1 & 1 & 1 & 0 & 0 & 0 \\
\mathbb{P}^1 & 0 & 0 & 0 & 1 & 1 & 0 & 0 \\
\mathbb{P}^2 & 1 & 0 & 0 & 0 & 1 & 1 & 0 \\
\mathbb{P}^1 & 0 & 0 & 1 & 0 & 0 & 0 & 1 \\
\mathbb{P}^1 & 0 & 0 & 0 & 0 & 0 & 0 & 2 \\
\mathbb{P}^1 & 0 & 0 & 0 & 0 & 0 & 0 & 2 \\
\mathbb{P}^1 & 0 & 0 & 0 & 0 & 0 & 1 & 1 \\
\end{array}
\right],
\end{eqnarray}
Our results show that this Calabi-Yau {\it does} have a freely acting symmetry descending from a linear action on this ambient space which leads to a smooth, non-simply connected, Calabi-Yau quotient. The symmetry group is $\mathbb{Z}_2$ and it acts on the homogeneous coordinates of the ambient space of $X_{6925 \text{\;favorable}}$ in the following manner.
 \begin{eqnarray} \label{conf6925gamma}
\gamma=\left(
\begin{array}{cccccccccccccccccc}
 1 & 0 & 0 & 0 & 0 & 0 & 0 & 0 & 0 & 0 & 0 & 0 & 0 & 0 & 0 & 0 & 0 & 0 \\
 0 & -1 & 0 & 0 & 0 & 0 & 0 & 0 & 0 & 0 & 0 & 0 & 0 & 0 & 0 & 0 & 0 & 0 \\
 0 & 0 & 0 & 0 & 0 & 0 & 0 & 1 & 0 & 0 & 0 & 0 & 0 & 0 & 0 & 0 & 0 & 0 \\
 0 & 0 & 0 & 0 & 0 & 0 & 0 & 0 & 1 & 0 & 0 & 0 & 0 & 0 & 0 & 0 & 0 & 0 \\
 0 & 0 & 0 & 0 & 0 & 0 & 0 & 0 & 0 & 1 & 0 & 0 & 0 & 0 & 0 & 0 & 0 & 0 \\
 0 & 0 & 0 & 0 & 0 & 1 & 0 & 0 & 0 & 0 & 0 & 0 & 0 & 0 & 0 & 0 & 0 & 0 \\
 0 & 0 & 0 & 0 & 0 & 0 & -1 & 0 & 0 & 0 & 0 & 0 & 0 & 0 & 0 & 0 & 0 & 0 \\
 0 & 0 & 1 & 0 & 0 & 0 & 0 & 0 & 0 & 0 & 0 & 0 & 0 & 0 & 0 & 0 & 0 & 0 \\
 0 & 0 & 0 & 1 & 0 & 0 & 0 & 0 & 0 & 0 & 0 & 0 & 0 & 0 & 0 & 0 & 0 & 0 \\
 0 & 0 & 0 & 0 & 1 & 0 & 0 & 0 & 0 & 0 & 0 & 0 & 0 & 0 & 0 & 0 & 0 & 0 \\
 0 & 0 & 0 & 0 & 0 & 0 & 0 & 0 & 0 & 0 & 0 & 0 & 0 & 0 & 0 & 0 & 1 & 0 \\
 0 & 0 & 0 & 0 & 0 & 0 & 0 & 0 & 0 & 0 & 0 & 0 & 0 & 0 & 0 & 0 & 0 & 1 \\
 0 & 0 & 0 & 0 & 0 & 0 & 0 & 0 & 0 & 0 & 0 & 0 & 1 & 0 & 0 & 0 & 0 & 0 \\
 0 & 0 & 0 & 0 & 0 & 0 & 0 & 0 & 0 & 0 & 0 & 0 & 0 & -1 & 0 & 0 & 0 & 0 \\
 0 & 0 & 0 & 0 & 0 & 0 & 0 & 0 & 0 & 0 & 0 & 0 & 0 & 0 & 1 & 0 & 0 & 0 \\
 0 & 0 & 0 & 0 & 0 & 0 & 0 & 0 & 0 & 0 & 0 & 0 & 0 & 0 & 0 & -1 & 0 & 0 \\
 0 & 0 & 0 & 0 & 0 & 0 & 0 & 0 & 0 & 0 & 1 & 0 & 0 & 0 & 0 & 0 & 0 & 0 \\
 0 & 0 & 0 & 0 & 0 & 0 & 0 & 0 & 0 & 0 & 0 & 1 & 0 & 0 & 0 & 0 & 0 & 0 \\
\end{array}
\right)
\end{eqnarray}
The symmetry also acts non-trivially on the defining polynomials as follows.
 \begin{eqnarray} \label{conf6925rho}
\gamma_p=\left(
\begin{array}{ccccccc}
 0 & 1 & 0 & 0 & 0 & 0 & 0 \\
 1 & 0 & 0 & 0 & 0 & 0 & 0 \\
 0 & 0 & 0 & 0 & 0 & 1 & 0 \\
 0 & 0 & 0 & 0 & 1 & 0 & 0 \\
 0 & 0 & 0 & 1 & 0 & 0 & 0 \\
 0 & 0 & 1 & 0 & 0 & 0 & 0 \\
 0 & 0 & 0 & 0 & 0 & 0 & 1 \\
\end{array}
\right).
\end{eqnarray}

Such results are not of course a surprise. It has been known since the classification was performed that the symmetries found in \cite{Braun:2010vc} were probably only a subset of those that exist on the Calabi-Yau threefolds being studied. Here we have simply provided a concrete description of some of the missing possibilities, concentrating on those that are associated to the most practically useful descriptions of the manifolds in question (favorable configuration matrices).

\section{Conclusions and Future Directions} \label{concsec}

The most commonly used construction of non-simply connected Calabi-Yau threefolds is to quotient a simply connected space by a freely acting discrete symmetry. The freely acting discrete symmetries that have been classified in the literature are described in terms of a linear action on some ambient space, which then induces the symmetry action on the Calabi-Yau manifold embedded inside it. Following such a methodology, different symmetries will be found by embedding the Calabi-Yau threefolds in different ambient spaces. In practical applications, the most useful descriptions of Calabi-Yau manifolds tends to be those in which the embedding inside an ambient space is `favorable' so that a complete understanding of the divisors on the variety can be obtained from restricting ambient space objects. In \cite{Braun:2010vc}, a classification was made of symmetries of the complete intersection Calabi-Yau threefolds in products of projective spaces which descend from the ambient spaces for these manifolds that were originally given in \cite{Candelas:1987kf}. In this paper we have classified a new set of cyclic symmetries of the same manifolds, descending from the ambient spaces of the favorable descriptions of these Calabi-Yau threefolds given in \cite{Anderson:2017aux}. In doing so, we have also presented a new class of non-simply connected Calabi-Yau manifolds.

We have presented a set of 129 symmetries. Of these, some are equivalent to the symmetries found in \cite{Braun:2010vc}, but phrased in terms of a different description of the Calabi-Yau threefold. Some are definitely new however. For example, 33 of these symmetries are of an order that simply was not known to exist on the relevant manifolds before this work. Our results are available, in a format described at the start of Section \ref{statres} at \href{http://www1.phys.vt.edu/~grayphys/CyclicSyms.html}{this link}.

\vspace{0.3cm}

In addition to providing new freely acting symmetries/non-simply connected Calabi-Yau manifolds, this work provides a rare insight into how much the obvious symmetries of a Calabi-Yau threefold depend upon the ambient space in which we embed it. By performing two different classifications for different choices of ambient spaces of the same manifolds, we gain some information as to how common it is to miss symmetries by restricting to one description. In the original CICY list whose symmetries were classified in \cite{Braun:2010vc} there are 2946 manifolds whose description is not favorable, and so for which the configuration matrix in the list provided in \cite{Anderson:2017aux} is different. For this set of manifolds our work finds 129 cyclic symmetries descending from the favorable ambient spaces whereas the results of \cite{Braun:2010vc} found 506 cyclic symmetries descending from the non-favorable ones. As mentioned above, at least 33 of the symmetries that we have found have an order that was not in the set classified in \cite{Braun:2010vc} and so are definitely new. Thus we see that, as was expected, the two sets of symmetries have significant differences and classifying symmetries that descend from one ambient space description does not give anything like a complete picture of the symmetries of a manifold. Indeed, if more descriptions of these same CICYs were studied one would expect to find  further symmetries beyond those discussed here. It is interesting to note that relatively fewer symmetries were found in the favorable CICY list than in the original description of these manifolds. This is the second time in the recent literature \cite{Braun:2017juz} that a search for non-simply connected manifolds has lead to relatively sparse results. It will be interesting to see if this trend continues in future searches and if this is an early indication that non-simply connected Calabi-Yau threefolds are rather rare. If true, this fact would be of great interest in string model building.

\vspace{0.2cm}

A first obvious extension of the work we present here would be to classify the non-cyclic symmetries descending from the ambient spaces of the favorable CICY descriptions of \cite{Anderson:2017aux}. One might also consider searching for symmetries descending from different favorable ambient spaces than the ones presented in that reference and in this regard, in Section \ref{labelforconc}, we gave a suggestion. As we discussed there, there are certain ways of generating favorable descriptions of manifolds which are more likely to make symmetries of certain orders manifest. It would be interesting to explore this further and see if it is possible to start with a type of symmetry you would like to have, and then tailor your description of the manifold to that symmetry (if it is indeed realized in the Calabi-Yau manifold). Finally, one could, of course, apply similar techniques to manifolds embedded in different ambient spaces than products of projective spaces, a study that has been initiated in \cite{Braun:2017juz}. Indeed, if sets of manifolds of the type introduced in \cite{Anderson:2015iia} (see \cite{Berglund:2016yqo,Berglund:2016nvh,Garbagnati:2017rtb,Jia:2018iza} for related work) could be classified, then one could even pursue similar investigations in cases where the Calabi-Yau manifold is described, not by a complete intersection, but by a generalized complete intersection.

\section*{Acknowledgements}

The authors would like to thank Lara Anderson, Matthew Brown, Wei Cui, and Weicheng Xue for helpful discussions. The work of J.G. is supported in part by NSF grant PHY-2014086. The authors acknowledge Advanced Research Computing at Virginia Tech for providing computational resources and technical support that have contributed to the results reported within this paper.

\appendix

\section{Appendix} \label{mrappendix}

In this appendix, we reproduce some well known results which are useful in carrying out the algorithm, first developed in \cite{Braun:2010vc}, that is described in the main text of the paper.

If one wants to compute the character of a $p$'th symmetric or anti-symmetric power of a representation $r$ which has character $\chi_r$ one can use the following recursive formulae \cite{fulton}.
\begin{eqnarray} 
\chi_{\textnormal{Sym}^p (r)} (g) &=& \frac{1}{p} \sum_{k=1}^p  \chi_r(g^k) \chi_{\textnormal{Sym}^{p-k}(r)}(g) \\ 
\chi_{\textnormal{Alt}^p(r)}(g) &=& \frac{1}{p} \sum_{k=1}^p (-1)^{k-1} \chi_r(g^k) \chi_{\textnormal{Alt}^{p-k}(r)}(g) 
\end{eqnarray}
For example, using these formulae we can obtain the following simple closed form expressions at low $p$.
\begin{eqnarray} \label{sym2}
\chi_{\textnormal{Sym}^2(r)}(g) &=& \frac{1}{2} \left( \chi_r(g)^2 + \chi_r(g^2)\right) \\ \label{alt2}
\chi_{\textnormal{Alt}^2(r)}(g) &=& \frac{1}{2} \left( \chi_r(g)^2 - \chi_r(g^2)\right) \\ 
\chi_{\textnormal{Sym}^3(r)}(g) &=& \frac{1}{6} \left( \chi_r(g)^3 + 3 \chi_r(g) \chi_r(g^2) + 2 \chi_r(g^3)\right)\\
\chi_{\textnormal{Alt}^3(r)}(g) &=& \frac{1}{6} \left( \chi_r(g)^3 - 3 \chi_r(g) \chi_r(g^2) + 2 \chi_r(g^3)\right)
\end{eqnarray}

The necessary formulae for $\text{Grind}$, that is $\text{SymInd}$ and $\text{AltInd}$, used in the main text were provided in \cite{Braun:2010vc}. We reproduce them here for convenience. If we are inducing from a subgroup $H$ to a group $G$ then the formulae only depend upon the index $[G:H]$ of the subgroup $H$ (recall that this quantity is related to the order of the two groups by $[G:H]= |G|/|H|$).
\begin{eqnarray}
\text{If}\;\; [G:H]=1:&&\\ \nonumber
\text{SymInd}_H^G (\chi_r) &=& \text{Ind}_H^G (\chi_r) = \chi_r \\ \nonumber
\text{AltInd}_H^G (\chi_r) &=& \text{Ind}_H^G(\chi_r) = \chi_r \\ \label{indwith2}
\text{If}\;\; [G:H]=2:&&\\ \nonumber
\text{SymInd}_H^G (\chi_r) &=& \text{Sym}^2 ( \text{Ind}_H^G(\chi_r) )- \text{Ind}_H^G( \text{Sym}^2(\chi_r)) \\ \nonumber
\text{AltInd}_H^G (\chi_r) &=& \text{Alt}^2 ( \text{Ind}_H^G(\chi_r) )- \text{Ind}_H^G( \text{Alt}^2(\chi_r))  \\
\text{If}\;\; [G:H]=3:&&\\ \nonumber
\text{SymInd}_H^G (\chi_r) &=& \text{Sym}^3(\text{Ind}_H^G (\chi_r) ) - \text{Ind}_H^G (\text{Sym}^3(\chi_r)) -\text{Ind}_H^G (\text{Sym}^2(\chi_r)) \text{Ind}_H^G (\chi_r) \\ \nonumber&& + \text{Ind}_H^G( \text{Sym}^{2,1}(\chi_r)) \\ \nonumber
\text{AltInd}_H^G (\chi_r) &=& \text{Alt}^3( \text{Ind}_H^G(\chi_r)) -\text{Ind}_H^G (\text{Alt}^3( \chi_r))  -\text{Ind}_H^G (\text{Alt}^2(\chi_r)) \text{Ind}_H^G (\chi_r) \\ \nonumber && + \text{Ind}_H^G( \text{Alt}^{2,1}(\chi_r))\\
\text{If}\;\; [G:H]=4:&&\\ \nonumber
\text{SymInd}_H^G (\chi_r) &=& \text{Sym}^4(\text{Ind}_H^G (\chi_r) ) - \text{Sym}^2( \text{Ind}_H^G(\text{Sym}^2(\chi_r))) - \text{Sym}^2( \text{Ind}_H^G(\chi_r)) \text{Ind}_H^G (\text{Sym}^2(\chi_r)) \\ \nonumber && +\text{Ind}_H^G ( \text{Sym}^{2,2}(\chi_r))-\text{Ind}_H^G ( \text{Sym}^{2,1,1}(\chi_r)) - \text{Ind}_H^G(\text{Sym}^3(\chi_r)) \text{Ind}_H^G(\chi_r) \\ \nonumber && + \text{Ind}_H^G ( \text{Sym}^{2,1}(\chi_r)) \text{Ind}_H^G(\chi_r) + \text{Ind}_H^G(\text{Sym}^2(\chi_r)) \text{Ind}_H^G (\text{Sym}^3(\chi_r))
 \\ \nonumber
\text{AltInd}_H^G (\chi_r) &=& \text{Alt}^4(\text{Ind}_H^G (\chi_r) ) + \text{Alt}^2( \text{Ind}_H^G(\text{Alt}^2(\chi_r))) - \text{Alt}^2( \text{Ind}_H^G(\chi_r)) \text{Ind}_H^G (\text{Alt}^2(\chi_r)) \\ \nonumber && +\text{Ind}_H^G ( \text{Alt}^{2,2}(\chi_r))-\text{Ind}_H^G ( \text{Alt}^{2,1,1}(\chi_r)) - \text{Ind}_H^G(\text{Alt}^3(\chi_r)) \text{Ind}_H^G(\chi_r) \\ \nonumber && + \text{Ind}_H^G ( \text{Alt}^{2,1}(\chi_r)) \text{Ind}_H^G(\chi_r) \\
\text{If}\;\; [G:H]=5:&&\\ \nonumber
\text{SymInd}_H^G (\chi_r) &=& \text{Sym}^5(\text{Ind}_H^G (\chi_r)) - \text{Ind}_H^G(\text{Sym}^5(\chi_r)) +2 \text{Sym}^4 (\text{Ind}_H^G(\chi_r)) \text{Ind}_H^G (\chi_r) \\ \nonumber && - \text{Sym}^2( \text{Ind}_H^G(\text{Sym}^2(\chi_r))) \text{Ind}_H^G(\chi_r) - \text{Sym}^3(\text{Ind}_H^G(\chi_r)) \text{Sym}^2(\text{Ind}_H^G(\chi_r)) \\ \nonumber && - \text{Sym}^3(\text{ind}_H^G(\chi_r)) \text{Ind}_H^G (\text{Sym}^2(\chi_r)) -9 \text{Ind}_H^G( \text{Sym}^{4,1}(\chi_r)) + \text{Ind}_H^G(\text{Sym}^{3,2}(\chi_r)) \\ \nonumber && +19 \text{Ind}_H^G(\text{Sym}^{3,1,1}(\chi_r)) - \text{Ind}_H^G(\text{Sym}^{2.2.1}(\chi_r)) -12 \text{Ind}_H^G( \text{Sym}^{3,1}(\chi_r)) \text{Ind}_H^G(\chi_r) \\ \nonumber && - 9 \text{Ind}_H^G(\text{Sym}^{2,1,1,1}(\chi_r) ) + \text{Ind}_H^G(\text{Sym}^{2,1,1}(\chi_r)) \text{Ind}_H^G(\chi_r)  \\ \nonumber && + 2 \text{Ind}_H^G (\text{Sym}^{1,1,1,1}(\chi_r)) \text{Ind}_H^G(\chi_r) + \text{Ind}_H^G(\text{Sym}^3(\chi_r)) \text{Ind}_H^G( \text{Sym}^2(\chi_r)) \\ \nonumber && +6 \text{Ind}_H^G ( \text{Sym}^{2,2}(\chi_r)) \text{Ind}_H^G(\chi_r)
\\ \nonumber
\text{AltInd}_H^G (\chi_r) &=& \text{Alt}^5(\text{Ind}_H^G (\chi_r)) - \text{Ind}_H^G(\text{Alt}^5(\chi_r)) +2 \text{Alt}^4 (\text{Ind}_H^G(\chi_r)) \text{Ind}_H^G (\chi_r) \\ \nonumber && + \text{Alt}^2( \text{Ind}_H^G(\text{Alt}^2(\chi_r))) \text{Ind}_H^G(\chi_r) - \text{Alt}^3(\text{Ind}_H^G(\chi_r)) \text{Alt}^2(\text{Ind}_H^G(\chi_r)) \\ \nonumber && - \text{Alt}^3(\text{Ind}_H^G(\chi_r)) \text{Ind}_H^G (\text{Alt}^2(\chi_r)) -9 \text{Ind}_H^G( \text{Alt}^{4,1}(\chi_r)) + \text{Ind}_H^G(\text{Alt}^{3,2}(\chi_r)) \\ \nonumber && +19 \text{Ind}_H^G(\text{Alt}^{3,1,1}(\chi_r)) +4\text{Ind}_H^G(\text{Alt}^{2.2.1}(\chi_r)) -12 \text{Ind}_H^G( \text{Alt}^{3,1}(\chi_r)) \text{Ind}_H^G(\chi_r) \\ \nonumber && - 9 \text{Ind}_H^G(\text{Alt}^{2,1,1,1}(\chi_r) ) + \text{Ind}_H^G(\text{Alt}^{2,1,1}(\chi_r)) \text{Ind}_H^G(\chi_r)  \\ \nonumber && + 2 \text{Ind}_H^G (\text{Alt}^{1,1,1,1}(\chi_r)) \text{Ind}_H^G(\chi_r) + \text{Ind}_H^G(\text{Alt}^3(\chi_r)) \text{Ind}_H^G( \text{Alt}^2(\chi_r)) 
\end{eqnarray}
In these expressions the short hand 
\begin{eqnarray}
\text{Sym}^{i_1,i_2,\ldots} (\chi_r)(g)= \prod_j \chi_{\text{Sym}^{i_j} (r)}(g)
\end{eqnarray}
has been used to denote relevant combinations of the symmetrization operations on characters given at the start of this appendix. Similarly the following has been defined for the anti-symmetric case.
\begin{eqnarray}
\text{Alt}^{i_1,i_2,\ldots} (\chi_r)(g)= \prod_j \chi_{\text{Alt}^{i_j} (r)}(g)
\end{eqnarray}

 %%%%%%%%%%%%%%%%%%%%%%%%%%%%%%%%%%%%%%%%%%%%%%%%%%%%%%%%%%%%%%%%%%%%%%%%%%%%%%%

%%%%%%%%%%%%%%%%%%%%%%%%%%%%%%%%%%%%%%%%%%%%%%%%%%%%%%%%%%%%%%%%%%%%%%%%%%%%%%%%%%%%%%%%%%%%%%%%

\end{document}